\documentclass[aps,prb,twocolumn,amssymb,amsfonts,groupedaddress,showpacs,epsf]{revtex4}
\usepackage{epsfig}
\usepackage{amsmath}
\usepackage{bm}
\usepackage{times}

\begin{document}
\title{Dynamical decoupling of a singlet-triplet qubit afflicted by a charge fluctuator}

\author{Guy Ramon}
\email{gramon@scu.edu}
\affiliation{Department of Physics, Santa Clara University, Santa Clara, CA 95053}

\begin{abstract}

The efficiency of dynamical decoupling pulse sequences in removing noise due to a nearby charge fluctuator is studied for a singlet-triplet spin qubit. We develop a numerical method to solve the dynamical equations for all three components of the Bloch vector under a general pulse protocol, where pulses are applied along an arbitrary rotation axis. The qubit is shown to undergo both dephasing and dissipative dynamics, pending on its working position. Analytical solutions are found for the limits of weakly and strongly coupled fluctuators, shedding light on the distinct dynamics in the different parameter regimes. Scaling of the qubit decay time with the number of control pulses is found to follow a power law over a wide range of parameters and qubit bias points.

\end{abstract}


\maketitle

\section{Introduction}

Considerable attention has been given in recent years to the proposal to encode the logical qubit states into spin singlet ($S$), and unpolarized triplet ($T_0$) states of two electrons localized in a double quantum dot (QD).\cite{Lidar,Levy,Taylor} In a Bloch sphere representation of the $S-T_0$ qubit, $|S\rangle =(|\!\!\uparrow\downarrow\rangle - |\!\!\downarrow \uparrow \rangle )/\sqrt{2}$ and $|T_0 \rangle=(|\!\!\uparrow\downarrow \rangle + |\!\!\downarrow \uparrow \rangle )/\sqrt{2}$ lie in the north and south poles, and rotations around the $z$ and $x$ axes are performed by the exchange interaction, $J$, and a magnetic field gradient across the two dots, $\delta h$, respectively. Concerted efforts have resulted in a series of impressive advances in the initialization, readout, control, and coupling of these systems. Rapid electrostatic control over the interdot bias provides a highly tunable exchange interaction that enables singlet preparation and single-shot readout by utilizing Pauli spin blockade, as well as fast rotations around the $z$ axis.\cite{Petta} Indeed, aside from their robustness against uniform nuclear fluctuations, the main advantage of $S-T_0$ qubits over single spin qubits is their amenability to fast single-qubit operations.

To complete single-qubit control, nuclear polarization cycles, in which bias is swept across the $S-T_+$ degeneracy point, have been used to exchange spin polarization between the electrons and the nuclei,\cite{PettaPRL,RamonNuc} and were shown to generate different Overhauser (hyperfine) fields in the two QDs.\cite{Foletti} Together with $J$, the resulting field gradient has provided an all-electric scheme to perform single qubit rotations around an arbitrary axis.\cite{Foletti} These nuclear pump cycles were later perfected by utilizing the hyperfine coupling in a feedback loop, which not only generated a stable nuclear field gradient of $\delta h=23$ mT, but also produced a narrowed nuclear state distribution, leading to a prolonged dephasing time of $T_2^*=94$ ns.\cite{BluhmPRL} Other methods to generate local magnetic field gradients were also demonstrated, including inhomogeneous Zeeman fields generated by on-chip micromagnets.\cite{Brunner}

Implementing two-qubit gates in a system of two double-dots has proven to be a challenging roadblock, and a first experimental demonstration of conditional operation on $S-T_0$ qubits was reported only last year.\cite{Weperen} Improved gate performance was recently demonstrated in a work that included a complete measurement of the system's density matrix using state tomography.\cite{Shulman} While both experiments used capacitive coupling to generate a CPHASE gate (that can be transformed into a CNOT gate with the addition of two Hadamard gates on the target qubit), the most notable new feature in the design of the latter experiment is the incorporation of nuclear state preparation that generated a stabilized field gradient. Shulman {\it et al.} then used this gradient ($\delta h =5$ mT, Zeeman energy of $0.125 \mu$eV) to apply a spin echo (SE) control pulse along the $x$ axis, mitigating charge noise and enabling them to work near the singlet avoided crossing, where sizable couplings between the two double dots can be obtained. At this bias, where $J \gg \delta h$, noise due to nuclear fluctuations is suppressed and charge noise plays a significant role. The employed SE pulse extended the two qubits' coherence time and allowed for their entanglement with a Bell state fidelity of 0.72.\cite{Shulman}

Evidently, dynamical decoupling (DD) pulse sequences have been applied to $S-T_0$ qubits prior to the above experiment with remarkable success, starting with a single pulse SE,\cite{Petta} and following by more advanced control sequences, including Carr-Purcell-Meiboom-Gill (CPMG), Concatenated DD (CDD), and Uhrig DD (UDD) schemes.\cite{Barthel,Bluhm,Medford} The coherence time was extended to a record of more than $200 \mu$s using a 16 pulse CPMG sequence and pulse optimization techniques.\cite{Bluhm} In these experiments, dephasing times were measured at a negative bias, where exchange is small, and decoherence is attributed mainly to the varying Overhauser fields. The $\pi$ pulses used in these sequences were thus appropriately performed (roughly) along the $z$ axis by pulsing the bias near the anticrossing where $J\gg \delta h$. The discussion above suggests, however, that when the qubit resides closer to the anticrossing (e.g., during two-qubit operations), nuclear fluctuations (causing $x$ rotations) will be suppressed and charge-noise-induced fluctuations (causing $z$ rotations) will become more prominent, thereby requiring $\pi_x$ pulses to  effectively extend the qubit coherence.

The effectiveness of various decoupling schemes at mitigating nuclear-induced dephasing has been extensively addressed theoretically for single-spin qubits\cite{Witzel1,Yao,Witzel2,Cywinski} and for two-spin qubits under SE.\cite{Neder} In contrast, their effectiveness in handling charge noise is less clear. In the current work we address this question by extending numerical and analytical stochastic methods that were developed in the context of superconducting qubits. We shall consider a qubit coupled to a single two-level-fluctuator (TLF), treating the latter as a classical source of random telegraph noise. While quantum telegraph noise was considered before (see, e.g., Ref.~\onlinecite{abel}), our classical treatment of the TLF seems reasonable, given the fact that the area surrounding the qubit is likely to be depleted of charge traps.

The qubit Hamiltonian reads ${\cal H}_q={\bf B} \cdot \mbox{\boldmath $\sigma$}$, where ${\bf B}=\frac{1}{2}(\delta h,0,J)$, and $\mbox{\boldmath $\sigma$}$ is the vector of Pauli spin matrices for the pseudospin states $S$ and $T_0$. Unless otherwise noted, we take $\delta h=0.125 \mu$eV, and the qubit working position, $\varphi=\arctan(\delta h/J)$, is determined by the interdot bias, $\varepsilon$, that controls $J$. Previous studies were mostly focused on the $\varphi =0$ point, where pure dephasing is expected, and the performance of the pulse sequence depends only on the TLF characteristics.\cite{Laikhtman,Paladino,Cywinski2} Bergli and Faoro considered periodic DD (PDD) at the $\varphi=\pi/2$ point, where both dephasing and dissipative dynamics take place.\cite{BerFao} Here we extend the work reported in Ref.~\onlinecite{BerFao}, by treating an arbitrary working position and a general pulse sequence.

While we present results pertaining for two-spin qubits in gate-defined GaAs double dots, our work is relevant for a variety of systems, including superconducting qubits and other QD materials. In particular, $S-T_0$ qubits were recently implemented in a Si/SiGe double QD, where dephasing time of 360 ns was measured.\cite{Maune} We expect that charge noise will play a dominant role in Si, where the hyperfine interaction strength is three orders of magnitude smaller, due to reduced coupling to- and number of nuclear spins, as compared with GaAs. DD has been successfully implemented in other systems such as electron spins in irradiated malonic acid single crystals.\cite{Du} Finally, very recently DD was incorporated with two-qubit gates in a hybrid system consisting of an electron spin and a nuclear spin in a single nitrogen-vacancy center in diamond.\cite{Sar}

The paper is organized as follows. In Sec.~II we present our model and discuss the qubit-TLF couplings and their dependence on various system parameters. In Sec.~III we detail the transfer matrix numerical approach to calculate qubit decoherence under a general pulse sequence. Sec.~IV provides analytical results for the PDD and CPMG cases, shedding light on the effectiveness of these sequences at different parameter regimes. In Sec.~V we use our formulation to estimate qubit coherence times for various scenarios, and in Sec.~VI we provide a short summary of our work, and outline future research directions. Appendix A includes formulae for the special case of pure dephasing, and Appendix B explains how to use the results given in Sec.~IV to explicitly find the dynamics of a qubit with an initial state along the equator of the Bloch sphere.

\section{qubit-TLF couplings}

A TLF residing near a double dot couples differently to the two-electron $S$ and $T_0$ spin states due to their different charge distributions.\cite{Hu} This results in fluctuations in the exchange interaction that lead to qubit decoherence and gate errors.\cite{Culcer,RamonTLF} Various sources can contribute to charge noise in lateral gated devices, including donor centers near the gate electrodes, switching events in the doping layer, and charge traps near quantum point contacts.\cite{Pioro,Taubert} Charge noise measurements in GaAs QDs revealed a linear temperature dependence characteristic of $1/f$ noise,\cite{Jung} which was shown theoretically to emerge from a TLF ensemble with an exponentially broad distribution of switching rates.\cite{Galperin,Schriefl,Bergli} These charge fluctuators behave classically and are characterized by switching rates, $\gamma_\pm$, and qubit coupling strength, $v$. In this picture we can write the qubit-TLF interaction as
\begin{equation}
{\cal H}_{\rm int} = v \xi (t) \sigma_z,
\end{equation}
where $\xi (t)=\pm 1$ is a classical noise representing a random telegraph process, switching between $\pm 1$ with rates $\gamma_\pm$.

In a previous work we have developed a multipole expansion technique to calculate the Coulomb couplings between the $S-T_0$ qubit orbital states and the fluctuator.\cite{RamonTLF} Assuming the TLF is a two-site trap, sufficiently remote from the double dot, there is no qubit-TLF tunnel coupling, and the interaction Hamiltonian includes the terms: $-v_\beta \sigma_z^Q -v_\gamma \sigma_z^Q \sigma_z^T$, where $\sigma_z^Q$ ($\sigma_z^T$) is the qubit (TLF) Pauli operator. We identify two possible scenarios for qubit-TLF couplings: (i) $\gamma$-coupled, where the charge fluctuates between two sites in the trap, and (ii) $\beta$-coupled, where the charge jumps in and out of the trap. $\beta$-coupled TLFs require a nearby charge reservoir (such as the 2DEG layer or quantum point contacts), and are expected to be less abundant. In the classical limit considered here, the qubit-TLF Hamiltonian reads:
\begin{equation}
{\cal H}={\cal H}_q+{\cal H}_{\rm int}={\bf B} (t) \cdot \mbox{\boldmath $\sigma$}, \label{Ham}
\end{equation}
where ${\bf B} (t)$ is
\begin{equation}
{\bf B}(t)= \left\{ \begin{array}{ll} \frac{1}{2}(\delta h,0,J-2v_\gamma \xi(t)), & \gamma-{\rm coupling} \\*[0.2 cm]
\frac{1}{2}(\delta h,0,J+v_\beta-v_\beta \xi(t)), & \beta -{\rm coupling} \end{array} \right.
\end{equation}
It is stressed that in the case of classical TLF considered in the current work, the two couplings have the same qualitative effect on the qubit dynamics, and here we distinct them only for the purpose of evaluating their strength.

The $v_\beta$ and  $v_\gamma$ couplings were calculated in Ref.~\onlinecite{RamonTLF} within a Hund-Mulliken orbital model using a multipole expansion to quadrupole-quadrupole order. It was shown that the leading term in the $\beta$-type ($\gamma$-type) coupling is dipole-charge (dipole-dipole), where left and right entries correspond to the qubit and TLF, respectively.\cite{RamonArx} Analytical approximations of these  couplings are found by keeping only a subset of the orbital two-electron states: $S(0,2),S(1,1)$, where $(i,j)$ indicate the number of electrons in each dot. Within this simplified model we find the leading terms:
\begin{eqnarray}
v_\beta^{21}(\varepsilon)\!&\!=\!&\! \frac{c \tilde{a}_q}{\tilde{R}^2} \frac{J^2 (\varepsilon)}{2J^2 (\varepsilon)+T_c^2} \sin \theta \cos \phi \label{b21} \\
v_\gamma^{22}(\varepsilon) \!&\!=\!&\! \frac{c\tilde{a}_q \tilde{a}_t}{\tilde{R}^3} \frac{J^2 (\varepsilon)}{2J^2 (\epsilon)+T_c^2} \left[\sin \theta_T \cos \phi_T -3\sin \theta \cos \phi \right. \notag \\ && \left. \times \sin \theta \sin \theta_T \cos(\phi-\phi_T)+\cos \theta \cos \theta_T \right], \label{g22}
\end{eqnarray}
where the left (right) superscript denotes contribution from a particular multipole moment of the qubit (TLF): charge (1), dipole (2), etc. Here, couplings are normalized to QD confinement energy, $\hbar \omega_0$, $c=(e^2/\kappa a_B)/\hbar \omega_0$ is the Coulomb to confinement energy ratio, $a_q$ ($a_t$) is the half interdot (TLF intersite) separation, $R$ is the qubit-TLF distance, and tilde denotes length normalized to the QD Bohr radius $a_B$. Furthermore, $J(\varepsilon)$ is the bias-dependent exchange and $T_c$ is the Coulomb-assisted tunnel coupling between the two dots. The angular dependence of $v_\beta$ and $v_\gamma$ is specified by four angles $(\theta,\phi,\theta_T,\phi_T)$, where the first two define the orientation of the qubit-TLF axis, and the last two define the TLF intersite axis.\cite{angles}

\begin{figure}[tb]
\epsfxsize=0.9\columnwidth
\vspace*{-0.1 cm}
\centerline{\epsffile{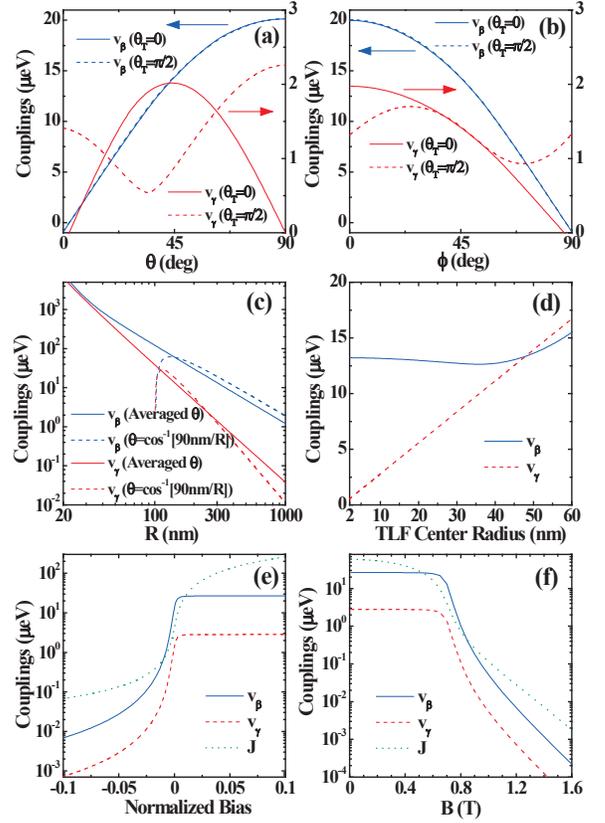}}
\vspace*{-0.1 cm}
\caption{(color online) qubit-TLF couplings vs. (a) polar angle $\theta$; (b) azimuthal angle $\phi$; (c) qubit-TLF distance $R$; (d) TLF Bohr radius $D_T$; (e) interdot bias normalized to QD confinement and measured from the singlet anticrossing point; (f) external magnetic field. In all plots blue (red) lines correspond to $v_\beta$ ($v_\gamma$). Left (right) $y$ axis in panels (a) and (b) correspond to $v_\beta$ ($v_\gamma$). Green lines in panels (e) and (f) depict the exchange interaction. Unless otherwise noted we consider the singlet anticrossing point (zero detuning), and qubit-TLF distance $R=300$ nm. Other system parameters are given in the main text.}
\label{Fig1}
\end{figure}

Fig.~\ref{Fig1} shows the dependence of the qubit-TLF couplings on various parameters. Here and throughout the paper we have diagonalized the full Hund-Mulliken Hamiltonian, which includes all two-electron Coulomb matrix elements. Nevertheless, the analytic formulas in Eqs.~(\ref{b21}), and (\ref{g22}) are instructive as they qualitatively capture many of the couplings features. We model the double dot with a quartic potential with dot confinement $\hbar \omega_0=3 {\rm meV}$ ($a_B\approx 20$ nm), and $a_q=2.8$, and use the dielectric constant for GaAs, $\kappa=13.1$. In addition, unless otherwise noted, we take the TLF center radius $D_t=5$ nm, and half intersite distance $a_t=20$ nm, chosen to characterize $\delta$-doped dopants in the insulator with a typical small radius and a large intersite separation. Lastly, unless otherwise noted, we consider an external magnetic field $B=0.7$ T applied along the device plane, perpendicular to the interdot axis, providing a Zeeman triplet splitting of $17.5 \mu$eV. This field has been used in recent experiments that utilized nuclear state preparation to generate Overhauser field gradients,\cite{BluhmPRL,Shulman} and in experiments that studied control sequences to enhance qubit coherence.\cite{Barthel,Medford}

In Figs.~\ref{Fig1}(a) and \ref{Fig1}(b) we examine the couplings dependence on the polar and azimuthal orientations of the qubit-TLF axis, respectively. We fix the TLF polar angle $\theta_T$ to either 0 (vertical traps) or $\pi/2$ (lateral traps) and look at the coupling dependence on $\theta, \phi$ for the two $\theta_T$ orientations, averaging out the remaining two angles. The blue lines in Figs.~\ref{Fig1}(a) and \ref{Fig1}(b) clearly replicate the angular dependence of $v_\beta^{21}$, given in Eq.~(\ref{b21}), with minimal dependence on $\theta_T$ from higher order terms. For $v_\gamma^{22}$, averaging over the remaining two angles results in
\[
\bar{v}_\gamma^{22} \sim  \sin 2\theta \cos \phi,
\]
for the case of $\theta_T=0$, in close agreement with the solid red lines in Figs.~\ref{Fig1}(a), and \ref{Fig1}(b). For $\theta_T=\pi/2$ we have
\begin{eqnarray*}
\bar{v}_\gamma^{22} (\theta) &\sim & \left|1-\frac{3}{2} \left( 1+\frac{2}{\pi} \right) \sin^2 \theta \right| \\
\bar{v}_\gamma^{22} (\phi) &\sim & \left| 1-3 (\cos 2\phi+ \sin 2\phi) \right|.
\end{eqnarray*}
Notice that at angles where $v_\gamma^{22}$ changes sign, contributions from higher order terms in the multipole expansion become important. In addition, the relative strength of these higher order terms increases with reduced qubit-TLF distance $R$, resulting in a more complicated angular dependence of the couplings. Both $v_\beta$ and $v_\gamma$ depend only mildly on the TLF azimuthal angle $\phi_T$ (not shown).

Fig.~\ref{Fig1}(c) shows the $R$ dependence of the couplings at zero detuning. Solid lines depict angular-averaged results, whereas dashed lines show couplings calculated by fixing the polar angle of the TLF-qubit axis to $\theta = \arccos (L_z/R)$, where $L_z=90$ nm is the depth of the 2DEG plane below the surface. We see that $v_\beta \sim R^{-2}$ for $R\gtrsim 50$ nm, and $v_\gamma \sim R^{-3}$ for $R\gtrsim 30$ nm, corresponding to the leading contributions in the multipole expansion. When $\theta$ is fixed by $R$, the leading terms in both $v_\beta$ and $v_\gamma$ change sign as $R$ approaches $L_z$ ($\theta \rightarrow 0$) and higher order terms become dominant. In the rest of the paper, having no a-priori knowledge of the TLF location and relative orientation, we calculate the couplings by performing averaging over all angles.\cite{angular}

In Fig.~\ref{Fig1}(d) we show the couplings dependence on the TLF site radius $D_t$, where the intersite half separation is varied accordingly as $a_t=4D_t$. $v_\gamma$ shows a linear dependence in $D_t$, matching the linear dependence of $v_\gamma^{22}$ on $a_t$. On the other hand, $v_\beta$ depends very mildly on $D_t$ as long as $D_t \lesssim 40$ nm, since its lowest order term that depends on $D_t$ and $a_t$ is dipole-quadrupole, which contributes only slightly to $v_\beta$ within the range of relevant trap sizes.
Fig.~\ref{Fig1}(e) shows $v_\gamma,v_\beta$, and $J$ dependence on the interdot bias point, $\varepsilon$,  normalized to QD confinement, where detuning is measured from the $S(1,1)-S(0,2)$  anticrossing point. The bias dependence is qualitatively approximated by the analytical formulas for the leading contributions, Eqs.~(\ref{b21}), and (\ref{g22}), with\cite{RamonDD}
\[
J(\varepsilon) \approx \frac{2 T_c^2}{\sqrt{\varepsilon^2+4 T_c^2}-\varepsilon}.
\]

Finally, the magnetic field dependence of $v_\gamma,v_\beta$, and $J$, is shown in Fig.~\ref{Fig1}(f). A decrease of more than four orders of magnitude in the couplings is observed between $B=100$ mT to $B=1.5$ T. We note that the magnetic compression of the orbitals reduces the wave function overlap by a factor of $\approx 10$ within this field range, thereby it alone cannot account for these results. In fact, the main reason for the high sensitivity of the couplings to the magnetic field is that varying $B$ also shifts the bias position of the singlet anticrossing. In Fig.~\ref{Fig1}(f) we have fixed the bias at the anticrossing point found for $B=0.7$ T. As $B$ increases, the anticrossing shifts to a more positive bias, and as a result our fixed bias becomes negative, thereby reducing the couplings. In experiments the bias position is likely calibrated when $B$ is varied and the high sensitivity of the exchange and qubit-TLF couplings would be greatly reduced.

We conclude that the qubit-TLF coupling strength may vary by several orders of magnitude and is particularly sensitive to the qubit working position (interdot bias) and the TLF distance. For concreteness, we consider in the rest of this paper only $\gamma$-type TLFs, given their relative abundance, and will henceforth drop the $\gamma$ subscript in $v_\gamma$ to avoid clutter.

\section{Transfer matrix method}
\label{seciii}

In this section we outline the formulation for qubit decoherence due to a single classical TLF.\cite{BerFao} We include asymmetric TLF switching, relevant in cases where the temperature is lower or comparable to the TLF level spacing.

We first write the qubit-TLF Hamiltonian, Eq.~(\ref{Ham}), in the basis of the qubit eigenstates:
\begin{equation}
{\cal H}=\Delta \sigma'_z+ v \xi(t)(\sin \varphi \sigma'_x -\cos \varphi \sigma'_z ), \label{Hfinal}
\end{equation}
where $\Delta=\frac{1}{2}\sqrt{\delta h^2+J^2}$, $\varphi=\arctan(\delta h/J)$, and $\sigma'_{x,z}$ are the Pauli matrices in the rotated frame. In this frame, the qubit evolves under a static field in the $z$ axis, with noise in both $x$ and $z$ axes. Since we are dealing with a bistable fluctuator, the Bloch vector representing the qubit state in the rotated frame precesses around either of the two effective fields
\begin{equation}
{\bf B}_\pm =(\pm v \sin \varphi, 0, \Delta \mp v \cos \varphi). \label{Bpm}
\end{equation}
We denote the two TLF states corresponding to ${\bf B}_+$ and ${\bf B}_-$ as up and down, respectively, and $\gamma_+$ ($\gamma_-$) is the switching rate from state up to down (down to up). Unlike the case of pure dephasing ($\varphi=0$), the Bloch vector does not precess around the $z$ axis alone but can reach any point on the Bloch sphere, resulting in dissipative dynamics. Denoting by $p({\bf r},t)$ the probability to reach point ${\bf r}=(x,y,z)$ on the Bloch sphere at time $t$, we introduce $p_+ ({\bf r},t)$ and $p_- ({\bf r},t)$ as the probabilities to reach the point ${\bf r}$ when the TLF is in state up and down, respectively.

The equations for $p_\pm ({\bf r},t)$ are found to be
\begin{eqnarray}
p_\pm({\bf r},t+\tau) \!&=&\!(1-\gamma_\pm \tau) p_\pm(U_\pm^{-1}{\bf r},t)+ \notag \\
&& \gamma_\mp \tau p_\mp(U_\pm^{-1}{\bf r},t),
\end{eqnarray}
where $U_\pm=e^{\tau {\bf B}_\pm \cdot {\bf R}}$ rotate the qubit around the ${\bf B}_\pm$ axes. We consider an infinitesimal time step $dt$, such that $\gamma_\pm dt$ are the switching probabilities for the two TLF states within time $dt$. Using the fields ${\bf B}_\pm$ in Eq.~(\ref{Bpm}), and the rotation matrices ${\bf R}\equiv (R_x,R_y,R_z),$\cite{BerFao} we find the master equations:
\begin{eqnarray}
\dot{p}_\pm \!\!&\!=\!&\!\! \left[ \left( \Delta \mp v \cos \varphi \!\right)\! (y \partial_x \!-\!x \partial_y) \pm v \sin \varphi (z \partial_y\!-\!y\partial_z) \right]\! p_\pm \notag \\
&& +(\gamma_\mp p_\mp -\gamma_\pm p_\pm). \label{ppm}
\end{eqnarray}
The expectation values of the Bloch vector components in the rotated frame evolving under the fields ${\bf B}_\pm$ are defined as:
\begin{equation}
{\bf r}_\pm=\int d{\bf r} p_\pm ({\bf r},t) {\bf r},
\end{equation}
and their dynamical equations are found from Eqs.~(\ref{ppm}) to be
\begin{eqnarray}
\dot{x}_\pm \!\!&\!=\!&\!\! (\gamma_\mp x_\mp -\gamma_\pm x_\pm) -\left(\Delta\mp v \cos \varphi \right) y_\pm \notag \\
\dot{y}_\pm \!\!&\!=\!&\!\! (\gamma_\mp y_\mp -\gamma_\pm y_\pm) +\left(\Delta\mp v \cos \varphi \right) x_\pm \mp v \sin \varphi z_\pm \notag \\
\dot{z}_\pm \!\!&\!=\!&\!\! (\gamma_\mp z_\mp -\gamma_\pm z_\pm) \pm v \sin \varphi y_\pm.
\end{eqnarray}
Finally we transform these coupled equations to a set of equations for the total expectation values, $x=x_+ +x_-, y=y_+ +y_-, z=z_+ +z_-$, and their differences,
$\delta x=x_+ -x_-, \delta y=y_+ -y_- ,\delta z=z_+ -z_-$:
\begin{eqnarray}
\dot{x} &=& -\Delta y+v \cos \varphi \delta y \notag \\
\dot{y} &=& \Delta x-v \cos \varphi \delta x-v \sin \varphi \delta z \notag \\
\dot{\delta z} &=& -\delta \gamma z -2 \gamma \delta z+v \sin \varphi y \notag \\
\dot{\delta x} &=& -\delta \gamma x-2 \gamma \delta x -\Delta \delta y+v \cos \varphi y \notag \\
\dot{\delta y} &=& -\delta \gamma y-2 \gamma \delta y +\Delta \delta x-v \cos \varphi x-v \sin \varphi z \notag \\
\dot{z} &=& v \sin \varphi \delta y, \label{xyz}
\end{eqnarray}
where we have defined the average switching rate, $\gamma=(\gamma_+ + \gamma_-)/2$, and switching rate difference, $\delta \gamma =\gamma_+ -\gamma_-$. It is convenient to write Eqs.~(\ref{xyz}) in a matrix form
\[
\dot{{\bf k}}=M_1 {\bf k},
\]
where entries 1,2,6 of the vector
\begin{equation}
 {\bf k} = \left( \begin{array}{c} x \\ y \\  \delta z \\ \delta x \\ \delta y \\ z \end{array} \right)
\end{equation}
define the Bloch vector components, and the matrix $M_1$ reads
\begin{equation}
M_1\!=\!\!\left(\!\!\!\! \begin{array}{cccccc} 0 \!&\! -\Delta \!&\! 0 \!&\! 0 \!&\! v \cos \varphi \!&\! 0 \\
\Delta \!&\! 0 \!&\! -v \sin \varphi \!&\! -v \cos \varphi \!&\! 0 \!&\! 0 \\
0 \!&\! v \sin \varphi \!&\! -2 \gamma \!&\! 0 \!&\! 0 \!&\! -\delta \gamma \\
-\delta \gamma \!&\! v \cos \varphi \!&\! 0 \!&\! -2 \gamma \!&\! -\Delta \!&\! 0 \\
-v \cos \varphi \!&\! -\delta \gamma \!&\! 0 \!&\! \Delta \!&\! -2 \gamma \!&\! -v \sin \varphi \\
0 \!&\! 0 \!&\! 0 \!&\! 0 \!&\! v \sin \varphi \!&\! 0 \end{array} \!\!\!\!\right)\!\!. \label{M1}
\end{equation}

We now consider the dynamics of the Bloch vector subject to sequences of control $\pi$ pulses along an arbitrary axis. The application of a $\pi$-pulse flips the components of the effective fields in Eqs.~(\ref{Bpm}) perpendicular to the pulse axis. While our dynamical equations are written in the rotated frame, we consider $\pi$-pulses along the original frame axes since these are more readily available in current experiments in $S-T_0$ qubits. \cite{Foletti} The resulting effective fields after the application of the control pulses read:
\begin{eqnarray}
{\bf B}_\pm^{\pi_x} \!&\!\!=\!\! &\! \left(\Delta \sin 2\varphi \mp v \sin \varphi,0,-\Delta \cos 2\varphi \pm v \cos \varphi \right) \notag\\
{\bf B}_\pm^{\pi_y}  \!&\!\!=\!\! &\! -\left( \pm v \sin \varphi,0,\Delta \mp v \cos \varphi \right) \notag\\
{\bf B}_\pm^{\pi_z}  \!&\!\!=\!\! &\!\left(-\Delta \sin 2\varphi \pm v \sin \varphi,0,\Delta \cos 2\varphi \mp v \cos \varphi \right)
\end{eqnarray}

\begin{widetext}
These fields lead to qubit state evolution governed by matrices $M_2^j$, where $j=x,y,z$ denotes the control pulse axis (in the original frame). $M_2^x$ is found as
\begin{equation}
M_2^x\!=\!\left(\!\! \begin{array}{cccccc} 0 \!&\! \Delta \cos 2\varphi \!&\! 0 \!&\! 0 \!&\! -v \cos \varphi \!&\! 0 \\ -\Delta \cos 2\varphi \!&\! 0 \!&\! v \sin \varphi \!&\! v \cos \varphi \!&\! 0 \!&\! -\Delta \sin 2\varphi \\ 0 \!&\! -v \sin \varphi \!&\! -2 \gamma \!&\! 0 \!&\! \Delta \sin 2\varphi \!&\! -\delta \gamma \\ -\delta \gamma \!&\! -v \cos \varphi \!&\! 0 \!&\! -2 \gamma \!&\! \Delta \cos 2\varphi \!&\! 0 \\ v \cos \varphi \!&\! -\delta \gamma \!&\! -\Delta \sin 2\varphi \!&\! -\Delta \cos 2\varphi \!&\! -2 \gamma \!&\! v \sin \varphi \\ 0 \!&\! \Delta \sin 2\varphi \!&\! 0 \!&\! 0 \!&\! -v \sin \varphi \!&\! 0 \end{array} \!\!\right)\!,
\end{equation}
$M_2^y$ is found by substituting $\Delta \rightarrow -\Delta$, $v \rightarrow -v$ in $M_1$, and $M_2^z$ is found by substituting $\Delta \rightarrow -\Delta$, $v \rightarrow -v$ in $M_2^x$.
\end{widetext}

The evolution of the qubit state after a sequence of $N$ $\pi_j$-pulses can be written as
\begin{equation}
{\bf k} (t)={\cal U}_j {\bf k}(0) \label{ft}
\end{equation}
with
\begin{equation}
{\cal U}_j= \left\{ \begin{array}{ll}\displaystyle{\prod_{i=1}^{\frac{1}{2}(N+1)}} e^{M_2^j \tau_{2i}} e^{M_1 \tau_{2i-1}}, &  N\in {\rm odd} \\[0.2 cm] e^{M_1 \tau_{N+1}} \displaystyle{\prod_{i=1}^{N/2}} e^{M_2^j \tau_{2i}} e^{M_1 \tau_{2i-1}}, &  N\in {\rm even} \end{array} \right. \label{T}
\end{equation}
where $\tau_i$ is the time interval between the $(i-1)$th and $i$th pulses.

When $\varphi =\pi/2$, and for symmetric switchings ($\delta \gamma =0$), the matrices $M_1$ and $M_2^y$ become block diagonal and the equations for $(x,y,\delta z)$ decouple from the equations for $(\delta x,\delta y,z)$.\cite{BerFao} Previously found results for pure dephasing ($\varphi =0$) can be obtained more easily by solving a subset of Eqs.~(\ref{xyz}), as shown in Appendix A.

For all the results presented below we take the initial state of the qubit to lie along the rotated $x$ axis, tilted by an angle $\varphi$ in the $x$-$z$ plane. Such initialization can be realized in two steps. First, the initial singlet state is transformed into the ground state of the nuclear field, $(|S \rangle +| T_0 \rangle )/\sqrt{2}$, by lowering the exchange adiabatically with respect to the nuclear mixing time.\cite{Petta} Second, a magnetic field gradient is ramped slowly so that the qubit state follows it, reaching the Hamiltonian eigenstate. We remark that this last step is challenging with current experimental techniques that generate nuclear field gradient, since the employed dynamical nuclear polarization cycles are slower than the electron spin dynamics. Our choice of initial state is motivated by the clean context it provides for the analysis of the qubit dynamics. In Appendix B, we provide formulas for qubit decay, when its initial state is set as $(|S \rangle +| T_0 \rangle )/\sqrt{2}$, and discuss its relation to the results given in the main text.

The probabilities of finding the TLF state up or down are $\gamma_-/(\gamma_+ +\gamma_-)$ and  $\gamma_+/(\gamma_+ +\gamma_-)$, respectively, therefore
\begin{equation}
p_\pm ({\bf r},t=0)=\frac{\gamma_\mp}{\gamma_+ +\gamma_-} \delta^{(3)} ({\bf r}-\hat{{\bf x}}),
\end{equation}
and the initial qubit's state reads
\begin{equation}
{\bf k}(0)= \left(1,0,0, -\frac{\delta \gamma}{2 \gamma},0,0 \right). \label{k0}
\end{equation}
The ratio between the TLF switching rates is given by the Boltzmann factor
\begin{equation}
\frac{\gamma_-}{\gamma_+}=e^{-\Delta E_t/k_B T},
\end{equation}
where $\Delta E_t$ is the TLF level splitting, taking into account intersite tunneling.\cite{RamonTLF}
For the TLF parameter range considered in this work, $\Delta E_t < 5\mu$eV, thus for sample temperature of 100 mK, we always have $\delta \gamma =2 \gamma \tanh (\Delta E_t/2k_B T) < \gamma$.

\begin{figure}[tb]
\epsfxsize=0.85\columnwidth
\vspace*{0 cm}
\centerline{\epsffile{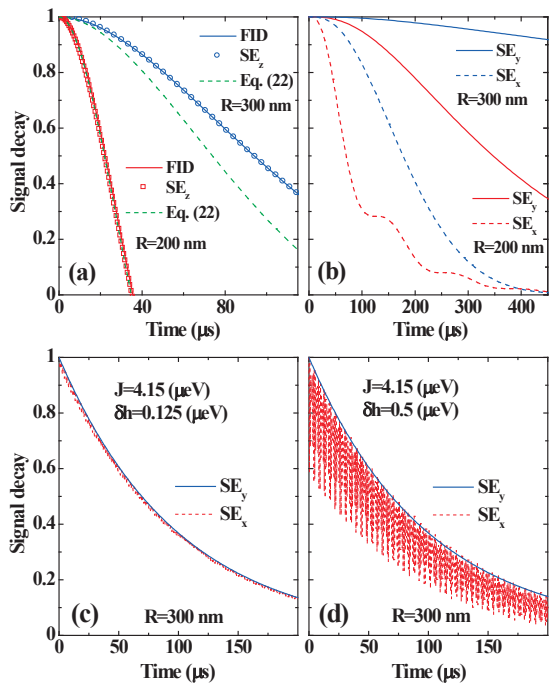}}
\vspace*{-0.2 cm}
\caption{(color online) Qubit decay under a SE control pulse with different rotation axes. (a) $\pi_z$ pulse (symbols) and FID (solid lines) decays for TLF distances of $R=200$ nm and $R=300$ nm. The qubit is at large negative bias, $J=0.02 \mu$eV, and $\delta h=0.125 \mu$eV ($\varphi \lesssim \pi/2$). Dashed green lines correspond to the strong coupling approximation, Eq.~(\ref{FID}); (b) qubit decay under $\pi_x$ and $\pi_y$ pulses for same parameters as in  panel (a); (c) qubit decay under $\pi_x$ and $\pi_y$ at zero detuning, where $J=4.15 \mu$eV $\gg \delta h=0.125 \mu$eV ($\varphi \approx 0$), for $R=300$ nm ($v=1.4 \mu$eV); (d) Same as panel (c) with $\delta h=0.5 \mu$eV. In all plots the TLF average switching time is 0.1 ms.}
\label{Fig2}
\end{figure}

We now address the choice of rotation axis for the control pulses. First, since charge noise is induced along the original $z$ axis, we expect that its decohering effects will not be corrected by $\pi_z$ pulses, which have been extensively employed to reduce nuclear induced noise. This is demonstrated in Fig.~\ref{Fig2}(a) where we plot qubit decay under both free induction (FID), and a single $\pi_z$ pulse (SE).\cite{coherent} Here we take TLF switching time of 0.1 ms and consider a large negative bias such that $J=0.02 \mu$eV. For a qubit-TLF distance, $R=200$ nm ($v=0.2$ neV), we have $v \gg \gamma$, where the FID decay is dominated by $v$:
\begin{equation}
\chi_{\rm FID} \approx e^{-\gamma t} \left( \cos vt+\frac{\gamma}{v} \sin v t \right). \label{FID}
\end{equation}
At $R=300$ nm, $v=62 $peV $\gtrsim \gamma$, and the strong coupling approximation is compromised [see dashed green lines in Fig.~\ref{Fig1}(a)]. In both cases the $\pi_z$ pulse does not extend the qubit coherence as compared with FID, resulting in dephasing times of $22 \mu $s, and $95 \mu$s, respectively. These relatively short dephasing times were likely circumvented in experiments by selecting relatively quiet samples.

The effectiveness of $\pi_x$ vs. $\pi_y$ control pulses in mitigating charge noise depends on the TLF parameters, as well as on the qubit working position. At pure dephasing ($\delta h=0$), both pulse types eliminate noise from TLFs that are static during the time interval between pulses, $\tau$, since the qubit state evolution depends only on the $\sigma_z$ operator [see Eq.~(\ref{Hfinal}), and Ref.~\onlinecite{BerFao} for more details]. At any other working position, $\pi_y$ pulses are always superior to $\pi_x$ pulses, where the relative improvement depends on the TLF parameters. The most pronounced improvement is expected at large negative bias where $\varphi \lesssim \pi/2$, as verified in Fig.~\ref{Fig2}(b). In Fig.~\ref{Fig2}(c) we tune the interdot bias to the singlet anticrossing (zero detuning), resulting in $J=4.15 \mu$eV $\gg \delta h$. Since $\varphi \approx 0$ the performance of $\pi_x$ and $\pi_y$ are similar. However if we stay at the anticrossing and take $\delta h = 0.5 \mu$eV, increased dephasing is exhibited by the $\pi_x$ pulse [see red dashed line in Fig.~\ref{Fig2}(d)]. Two-qubit gates that are performed near zero detuning (to increase coupling strength) will thus benefit greatly from $\pi_y$ pulses that will allow to increase $\delta h$, shortening gate times.

Finally we note that it is possible to effectively eliminate the exchange interaction so that$\varphi = \pi/2$ while retaining substantial qubit-TLF coupling. Normally, $J$ and $v$ decrease simultaneously with bias [see Fig.~\ref{Fig1}(e)], but, as was recently shown, at certain interdot bias points, the inter-qubit Coulomb couplings balance the internal exchange interaction, resulting in a zero effective exchange.\cite{RamonDD} For a distance of 400 nm between the centers of the two DDs, this zero-exchange position is found at a moderately negative (dimensionless) bias of $\varepsilon=-0.05$. While this position is typically ideal from the perspective of charge noise, for sufficiently close TLFs, $\pi_x$ pulses will introduce substantial noise.

The above considerations suggest that $\pi_y$ pulses are more effective in mitigating charge noise in several important scenarios. Moreover, we expect $\pi_x$ pulses to be completely inefficient in correcting nuclear-induced noise, similarly to the inadequacy of $\pi_z$ pulses in correcting charge noise. This will play an important role at negative bias, where nuclear noise is dominant. While $\pi_y$ pulses are harder to implement and were previously generated with limited fidelity,\cite{Foletti} it is expected that high-fidelity $\pi_y$ pulses will be available in the near future.\cite{Hendrik}

\section{Analytic results for PDD and CPMG pulse sequences}
\label{anal}

In this section we provide exact analytic solutions for the qubit dynamics due to its coupling with a single TLF, in the limits of weakly ($v \ll \gamma$) and strongly ($v \gg \gamma$) coupled fluctuators. As we show below, the qubit dynamics is different in these two regimes, and the asymptotic behavior in these limits proves valuable for the interpretation of our results. Our analysis follows Ref.~\onlinecite{BerFao}, extending it in three respects: (i) we consider an arbitrary qubit working point $\varphi$, (ii) we include asymmetric TLF switching, and (iii) we find explicitly the weights of the different decay rates in the final solution, providing further insight into the qubit dynamics, and the enhanced performance of CPMG over PDD control pulses.

In the following we consider an $N$-pulse PDD protocol with a constant time interval between pulses, $\tau=t/(N+1)$, where $t$ is the total time. (CPMG protocol will be considered in section \ref{secCPMG}.) In light of the discussion ending the previous section, we consider only $\pi_y$ pulses. In order to calculate the qubit dynamics, we first note that after a $\pi_y$ pulse is applied, the matrix that governs the qubit evolution, $M_2^y$, can be written in terms of $M_1$ in Eq.~(\ref{M1}) as:
\begin{equation}
M_2^y=L M_1 L,
\end{equation}
where $L={\rm diag}(1,-1,1,1,-1,1)$. Thus the qubit evolution under the full control sequence is found by
\begin{equation}
{\bf k}(t)= T^{N+1} {\bf k}(0), \label{kt}
\end{equation}
where ($N$ is odd for PDD)
\begin{equation}
 T=\sqrt{e^{M_2^y \tau} e^{M_1 \tau}} = L e^{M_1 \tau}. \label{Tanal}
\end{equation}
The decay rates of the Bloch vector components are found in terms of the six eigenvalues, $\chi_i$, of the evolution operator $T$:
\begin{equation}
\Gamma_i=-\frac{\ln |\chi_i|}{\tau}, \label{eta}
\end{equation}
and the  general solution for the decay of the qubit state rotated components is
\begin{equation}
j(t)=\sum_{i=1}^6 w_i^j e^{-\Gamma_i t}, \hspace{0.5 cm} j=x,y,z, \label{jt}
\end{equation}
where $w_i^j$ is the weight of the $i$th eigenvalue in the solution of the $j$th component. It should be noted that the decay rates defined in Eq.~(\ref{eta}) and calculated below, are time dependent, unlike the more commonly encountered decay coefficients obtained in Bloch-Redfield theory. Notice also that while we are only interested in the dynamics of the three rotated Bloch vector components, in the strong coupling case discussed below there are nonzero weights for all six eigenvalues of matrix $T$.

\subsection{PDD: Weak Fluctuator}

In order to find the eigenvalues $\chi_i$ of $T$ we need to exponentiate $M_1$ in Eq.~(\ref{M1}). In the case of weak coupling we perform second order (degenerate) perturbation in $v/\gamma$. Note that while $M_1$ is non hermitian due to the unequal switching rates, $\delta \gamma \neq 0$, we still have $M_1=M_1^R +iM_1^I$, where $M_1^R,M_1^I$ are both hermitian. In this case, the eigenstates of $M_1$ and $M_1^\dagger$ form a bi-orthogonal set and a perturbation theory is readily available.\cite{PRC} The three relevant decay rates are found as:
\begin{eqnarray}
\Gamma_1^w \!&\!=\!&\! \!\left(\! 1\!-\!\frac{\delta \gamma^2}{4\gamma^2}\right)\! \frac{\gamma v^2}{\Delta^2 +4 \gamma^2} \left[ (1\!-\!A\!-\!B_1)\sin^2 \varphi +C \cos^2 \varphi \right] \notag \\
\Gamma_{2,3}^w \!&\!=\!&\! \!\left(\! 1\!-\!\frac{\delta \gamma^2}{4\gamma^2}\right)\! \frac{\gamma v^2}{\Delta^2 +4 \gamma^2} \left[ (2\!-\!B_1\!-\!B_2)\sin^2 \varphi-F \right. \notag \\
&& \left. \mp \sqrt{F^2+D^2 \sin^2 2\varphi} \right], \label{G123w}
\end{eqnarray}
where the different functions of $\tau$ are given by:
\begin{eqnarray}
A\!&\!=\!&\!\frac{\Delta^2-4 \gamma^2}{\Delta^2+4 \gamma^2} {\rm sinc} \tilde{\tau} \notag \\
B_1 \!&\!=\!&\! \frac{8 \gamma^2}{\Delta^2 +4\gamma^2} \cos^2 \frac{\tilde{\tau}}{2} \frac{\tanh \gamma \tau}{\gamma \tau} \notag \\
B_2 \!&\!=\!&\! \frac{8 \gamma^2}{\Delta^2 +4\gamma^2} \sin^2 \frac{\tilde{\tau}}{2} \frac{\coth \gamma \tau}{\gamma \tau} \notag \\
C\!&\!=\!&\! \frac{\Delta^2+4 \gamma^2}{2\gamma^2} \left(1- \frac{\tanh \gamma \tau}{\gamma \tau} \right) \notag \\
D\!&\!=\!&\! \cos \frac{\tilde{\tau}}{2} \frac{\tanh \gamma \tau}{\gamma \tau} -{\rm sinc} \frac{\tilde{\tau}}{2} \notag  \\
F\!&\!=\!&\! \frac{1}{2} \left[ (1-A-B_1) \sin^2 \varphi -C \cos^2 \varphi \right]. \label{A-F}
\end{eqnarray}
Here we used ${\rm sinc}\tilde{\tau}\equiv\sin \tilde{\tau}/\tilde{\tau}$, where $\tilde{\tau} \equiv \Delta \tau$ is the normalized time interval between pulses. We note that the effect of asymmetric TLF switching in this regime amounts to reducing all rates by a common factor that becomes appreciable for TLF level splitting above $1\mu eV$. The three additional eigenvalues, associated with the dynamics of $(\delta x, \delta y, \delta z)$, induce much faster decay rates, $2 \gamma -O(v^2/\gamma^2)$, but their weights vanish for all three Bloch vector components. Eqs.~(\ref{G123w}), (\ref{A-F}) recover previous results obtained for $\varphi=0$ and $\varphi=\pi/2$ with $\delta \gamma =0$.\cite{BerFao} Specifically, for the case of pure dephasing ($\varphi =0$), we have a single decay rate (see also Appendix A):
\begin{equation}
\Gamma_1^w(\varphi =0)=\Gamma_3^w(\varphi =0)=\frac{v^2}{2\gamma} \left(1-\frac{\tanh\gamma \tau}{\gamma \tau} \right). \label{G1w}
\end{equation}
For $J=0$ ($\varphi =\pi/2$), the equations for $x$ and $y$ decouple from $z$, the latter having no dynamics when the qubit state is initially along the $x$ axis, and we obtain two decay rates:
\begin{eqnarray}
\Gamma_1^w (\varphi=\pi/2) \!&\!=\!&\! \! \frac{\gamma v^2}{\Delta^2 +4 \gamma^2} (1\!-\!A\!-\!B_1) \notag \\
\Gamma_2^w (\varphi=\pi/2) \!&\!=\!&\! \! \frac{\gamma v^2}{\Delta^2 +4 \gamma^2} (1\!+\!A\!-\!B_2). \label{G12w}
\end{eqnarray}

In order to explicitly find the time dependence of the Bloch vector components, we diagonalize the matrix $T$ in Eq.~(\ref{kt}), $T=SWS^\dagger$, where $W$ is diagonal matrix, similar to $T$, whose $i$th element is $e^{-\Gamma_i \tau}$, and the columns of $S$ are the eigenvectors of $T$. Up to second order in $v/\gamma$ we can use the unperturbed eigenvectors, and by applying the resulting operator to the initial vector, Eq.~(\ref{k0}), we find the following weights
\begin{eqnarray}
w_1^x \!&\!=\!&\! \sin^2 \frac{\tilde{\tau}}{2} \notag \\[-0.05 cm]
w_{2,3}^x \!&\!=\!&\! \frac{1}{2} \cos^2 \frac{\tilde{\tau}}{2} \left(1 \pm \frac{F}{\sqrt{F^2+D^2 \sin^2 2 \varphi}} \right)\notag \\[-0.05 cm]
w_1^y \!&\!=\!&\! \frac{1}{2} \sin \tilde{\tau} \notag \\[-0.05 cm]
w_{2,3}^y \!&\!=\!&\! -\frac{1}{4} \sin \tilde{\tau} \left(1 \pm \frac{F}{\sqrt{F^2+D^2 \sin^2 2 \varphi}} \right) \notag  \\
w_1^z \!&\!=\!&\! 0 \notag \\
w_{2,3}^z \!&\!=\!&\! \pm \frac{1}{2}\cos \frac{\tilde{\tau}}{2} \frac{D \sin 2 \varphi}{\sqrt{F^2+D^2 \sin^2 2 \varphi}}. \label{wijw}
\end{eqnarray}
Notice that $\sum_i w_i^x=1$, and $\sum_i w_i^y=\sum_i w_i^z=0$.

\begin{figure}[t]
\epsfxsize=0.82\columnwidth
\vspace*{-0.2 cm}
\centerline{\epsffile{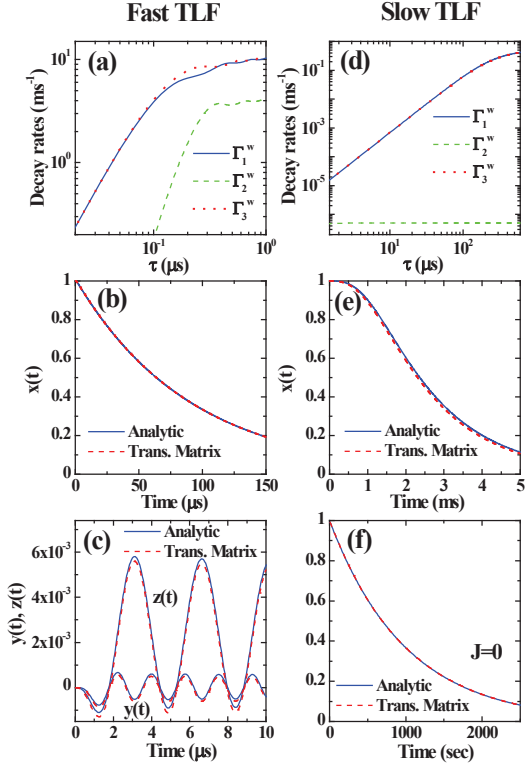}}
\vspace*{-0.2 cm}
\caption{(color online) Panels (a)-(c): Weakly-coupled fast TLF ($1/\gamma=0.1 \mu s$, $v=2.5$ neV), $\Delta \sim \gamma \gg v$. (a) Decay rates vs.~time interval between pulses, $\tau$, from Eqs.~(\ref{G123w}) and (\ref{A-F}); (b), (c) Bloch vector components under 11-pulse PDD sequence. Panels (d)-(f): Weakly-coupled slow TLF ($1/\gamma=0.1$ ms, $v=8.4$ peV), $\Delta \gg \gamma \gg v$. (d) Decay rates from Eqs.~(\ref{G123w}) and (\ref{A-F}); (e) $x$ component under 11-pulse PDD sequence ($y$ and $z$ components are almost unaffected by the TLF); (f) Same as panel (e) with $J=0$. In panels (b), (c), (e), and (f), solid blue lines depict Eqs.~(\ref{jt}) and (\ref{wijw}), and dashed red lines show transfer matrix calculation. In all plots $J=\delta h=0.125 \mu$eV ($\varphi=\pi/4$), except for panel (f) where $J=0$.}
\label{Fig3}
\end{figure}
In Fig.~\ref{Fig3} we illustrate the above results for a working position $\varphi =\pi/4$ ($J=\delta h =0.125 \mu$eV), obtained at negative bias $\epsilon=-0.054$. In Fig.~\ref{Fig3}(a) we consider a fast TLF with average switching time of $0.1 \mu$s, and coupling strength $v=2.5$ neV ($R=300$ nm), such that $\Delta \sim \gamma \gg v$. All three decay rates in Eqs.~(\ref{G123w}) are comparable and contribute to the decay of all three Bloch vector components shown in Figs.~\ref{Fig3}(b) and (c), where we consider the qubit dynamics under an 11-pulse PDD protocol. In Figs.~\ref{Fig3}(d)-(f) we consider a slow TLF with an average switching time of 0.1 ms. To stay in the weak coupling regime, we take $R=2000$ nm, resulting in $v=8.4$ peV, so that $\Delta \gg \gamma \gg v$. In this case the function $C$ in Eqs.~(\ref{A-F}) dominates and we obtain two distinct decay rates: $\Gamma^w_1 \approx \Gamma_3^w \approx \frac{v^2}{2 \gamma} (1-\frac{\tanh \gamma \tau}{\gamma \tau}) \cos^2 \varphi$, and $\Gamma_2^w \approx \frac{2 \gamma v^2}{\Delta^2} \sin^2 \varphi$, the latter being independent of $\tau$. Whereas $\Gamma^w_1,\Gamma^w_3 \gg \Gamma^w_2$, the weight $w_2^x$ is vanishingly small, and overall the qubit state decays with a single (faster) rate. We also observe that in this limit there is almost no dissipative dynamics since $w_2^y \approx 0$, and $w_1^y=-w_3^y$ for the $y$ component, and $w_{2,3}^z \sim (\gamma/\Delta)^2 \ll 1$ for the $z$ component. Finally, we note that the leading term in the faster decay rate, $\Gamma_1^w$, is eliminated at $\varphi =\pi/2$, and the qubit decay is governed by the much slower rate $\Gamma_2^w$, as well as by higher order corrections in $\gamma/\Delta$, and $v/\gamma$ to $\Gamma_1^w$ and $\Gamma_3^w$. This case is shown in Fig.~\ref{Fig3}(f), where we set $J=0$ and obtain more than 5 orders-of-magnitude increase in coherence time.

\begin{figure}[t]
\epsfxsize=0.87\columnwidth
\vspace*{0 cm}
\centerline{\epsffile{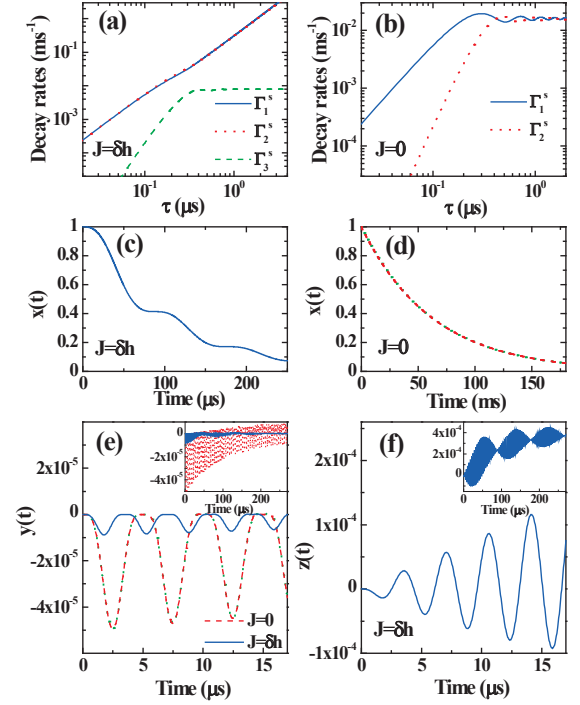}}
\vspace*{-0.2 cm}
\caption{(color online) Strongly coupled TLF with $v=2.5$ neV ($R=300$ nm) and $1/\gamma=0.1$ ms, for $\delta h=0.125 \mu$eV, satisfying $\Delta \gg v \gg \gamma$. (a), (b) Decay rates vs. time interval between pulses, $\tau$, from Eqs.~(\ref{G123s}) for $J=\delta h$ ($\varphi=\pi/4$), and $J=0$ ($\varphi=\pi/2$), respectively; (c), (d) qubit decay under 11-pulse PDD sequence for $J=\delta h$ and $J=0$, respectively; (e) Bloch vector $y$ component dynamics under 11-pulse PDD sequence (solid blue line) and $J=0$ (dashed red line); (f) Bloch vector $z$ component dynamics under 11-pulse PDD sequence for $J=\delta h$ (no $z$ dynamics for $J=0$). The dotted green lines in panels (d) and (e) correspond to Eqs.~(\ref{wijsx}) and (\ref{wijsy}). The insets in panels (e) and (f) depict the long time dynamics of the transverse Bloch vector components.}
\label{Fig4}
\end{figure}

\subsection{PDD: Strong Fluctuator}

\begin{figure}[tb]
\epsfxsize=0.9\columnwidth
\vspace*{0 cm}
\centerline{\epsffile{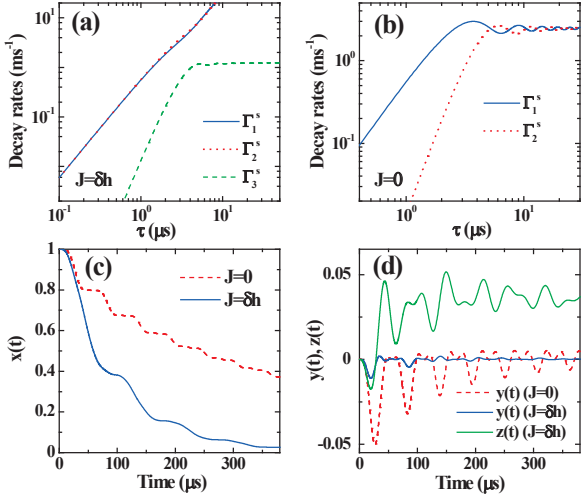}}
\vspace*{-0.1 cm}
\caption{(color online) Strongly coupled TLF with $v=2.5$ neV ($R=300$ nm) and $1/\gamma=0.1$ ms, for $\delta h=0.01 \mu$eV, satisfying $\Delta \gtrsim v \gg \gamma$. (a), (b) Decay rates vs. time interval between pulses, $\tau$, from Eqs.~(\ref{G123s}) for $J=\delta h$ ($\varphi=\pi/4$), and $J=0$ ($\varphi=\pi/2$), respectively; (c) qubit decay under 11-pulse PDD sequence for $J=\delta h$ and $J=0$; (d) $y$ and $z$ components dynamics under 11-pulse PDD sequence (no $z$ dynamics for $J=0$). Notice that for comparison with Fig.~\ref{Fig4} we kept the same coupling strength, although normally it would reduce with $J$. One can tune $J$ without changing the bias (thus keeping $v$ fixed) by introducing a second double dot.}
\label{Fig5}
\end{figure}

In the limit of strong coupling, $\gamma \ll v$, it is sufficient to perform first order perturbation in $\gamma/v$. Here the unperturbed eigenvalues of $T$ form two degenerate subspaces of ${\rm dim}=2$, and ${\rm dim}=4$. The analytic results are complicated and we present here only the leading term (third order) in the expansion in $v/\Delta$, in the case of symmetric switching ($\delta \gamma =0$). The results below are thus valid at negative detunings, where $v \ll \Delta$ (for a fixed $\delta h=0.125 \mu eV$), but will be applicable to an extended bias regime if larger $\delta h$ is considered. The three decay rates associated predominantly with the Bloch vector components are found as:

\begin{eqnarray}
\Gamma_1^s \!&\!=\!&\! \! \frac{\gamma v^2}{\Delta^2} \!\left( \tilde{A} \sin^2 \varphi +\frac{\tilde{\tau}^2}{6} \cos^2 \varphi \right) \notag \\
\Gamma^s_{2,3} \!&\!=\!&\! \! \frac{\gamma v^2}{\Delta^2} \!\left(\! \tilde{B} \sin^2 \varphi-\tilde{F} \pm \sqrt{\tilde{F}^2+\tilde{D}^2 \sin^2 2\varphi} \right)\!\!, \label{G123s}
\end{eqnarray}
where the different functions are given by
\begin{eqnarray}
\tilde{A}\!&\!=\!&\! 1- {\rm sinc} \tilde{\tau} \notag \\
\tilde{B} \!&\!=\!&\! 2 \left( 1-{\rm sinc}^2 \frac{\tilde{\tau}}{2} \right) \notag \\[-0.05 cm]
\tilde{D}\!&\!=\!&\! \cos \frac{\tilde{\tau}}{2} -{\rm sinc} \frac{\tilde{\tau}}{2} \notag \\[-0.05 cm] \tilde{F}\!&\!=\!&\! \frac{1}{2} \left( \tilde{A} \sin^2 \varphi -\frac{\tilde{\tau}^2}{6} \cos^2 \varphi \right). \label{A-C}
\end{eqnarray}
We consider strongly coupled TLF in two scenarios: $\Delta \gg v \gg \gamma$, and $\Delta \gtrsim v \gg \gamma$, shown in Figs.~\ref{Fig4} and \ref{Fig5}, respectively.

Examining the asymptotic behavior of the decay rates at $\tilde{\tau} \ll 1$, we have $\tilde{A}=\tilde{B}=-2\tilde{D}=\tilde{\tau}^2/6$. As a result, at short time intervals, the qubit dynamics is governed by a single decay rate: $\Gamma_1^s \approx \Gamma_2^s \approx \gamma v^2 \tau^2/6$, depicted by the solid blue and dotted red lines in Fig.~\ref{Fig4}(a). $\Gamma_3^s$  scales as $\tau^4$ and is considerably smaller. At the limit of $\tilde{\tau} \gg 1$ we have $\tilde{A},\tilde{B},\tilde{D} \ll \tilde{\tau}$, thus $\Gamma^s_1\approx \Gamma^s_2 \approx \gamma v^2 \tau^2 \cos^2 \varphi/6 \gg \Gamma^s_3 \approx 2 \gamma v^2 \sin \varphi^2/\Delta^2$. Note that for pure dephasing, both asymptotes coincide and $\Gamma^s (\varphi =0)= \gamma v^2 \tau^2/6$ is the single solution to Eqs.~(\ref{G123s}). We conclude that when $v \ll \Delta$, unless $\varphi=\pi/2$, the qubit dynamics is governed at all times by a single decay rate that scales as $\tau^2$. In contrast, when $J=0$, the $\cos^2 \varphi$ term is removed and $\Gamma^s_{1,2}$ decrease substantially in the large $\tilde{\tau}$ limit, becoming comparable to $\Gamma^s_3$. In this case the rates read:
\begin{eqnarray}
\Gamma_1^s (\varphi=\pi/2) \!&\!=\!&\! \frac{\gamma v^2}{\Delta^2} (1-{\rm sinc} \tilde{\tau}) \notag \\
\Gamma_2^s (\varphi=\pi/2) \!&\!=\!&\! \frac{\gamma v^2}{\Delta^2} \left(1-2{\rm sinc}^2 \frac{\tilde{\tau}}{2} +{\rm sinc} \tilde{\tau}\right), \label{G13}
\end{eqnarray}
both contribute to the qubit decay, and their magnitude in the large $\tilde{\tau}$ limit is curbed at a value $\gamma v^2/\Delta^2$, as shown in Fig.~\ref{Fig4}(b). As a result, decay time is nearly three orders of magnitude longer, as compared with the $J=\delta h$ case. This is shown in Figs.~\ref{Fig4}(c), and (d) for an 11 pulse PDD sequence. We note that in this case, the equations for $(x,y,\delta z)$ decouple from those for $(\delta x,\delta y,z)$, and $\Gamma^s_3$ is associated solely with the $z$ component, having no weight in the decay of our initial state.

In order to fully capture the qubit dynamics, we calculate the weights of the different rates in Eq.~(\ref{G123s}). As in the weak coupling case, the three additional eigenvalues, associated predominantly with the dynamics of $(\delta x, \delta y, \delta z)$, induce much faster decay rates: $2 \gamma -O(\gamma v^2/\Delta^2)$, but in this case they have a small yet nonzero weight that contributes to the decay of the Bloch vector. We demonstrate this behavior for $\varphi =\pi/2$, where the analytic formulas are simpler due to the block diagonal form of $M_1$ in Eq.~(\ref{M1}). Here we only need to consider the three eigenvalues of the block $(x,y,\delta z)$, where two decay rates are given in Eqs.~(\ref{G13}), and the third rate is $\Gamma_3 \approx 2\gamma$, associated predominantly with $\delta z$. Considering the perturbed eigenvectors of $T$ to second order in $v/\Delta$, the weights for the $x$ and $y$ components ($z$ has no dynamics) are calculates to be:
\begin{eqnarray}
w_1^x \!&\!=\!&\! \sin^2 \frac{\tilde{\tau}}{2} -\frac{(v\tilde{\tau})^2}{4\Delta^2} \left( \tilde{A}-\frac{\tilde{B}}{2} \right) \notag \\
w_2^x \!&\!=\!&\! \cos^2 \frac{\tilde{\tau}}{2}+ \frac{v^2}{\Delta^2} \left[ \frac{\tilde{\tau^2}}{4} \left(\tilde{A} -\frac{\tilde{B}}{2} \right)-\tilde{A}^2\right] \notag \\
w_3^x \!&\!=\!&\! \frac{v^2}{\Delta^2} \tilde{A}^2 \label{wijsx}
\end{eqnarray}
and
\begin{eqnarray}
w_1^y \!&\!=\!&\! \frac{1}{2} \sin \tilde{\tau} +\frac{v^2\tilde{\tau}}{4\Delta^2} \left[\tilde{A}+ \frac{\tilde{\tau}^2}{4} (\tilde{B}-2) \right] \notag \\
w_2^y \!&\!=\!&\! -\!\frac{1}{2} \sin \tilde{\tau}- \frac{v^2\tilde{\tau}}{4\Delta^2} \left[(\tilde{B} -2) \left( \frac{\tilde{\tau}^2}{4} -\tilde{A} \right)+\tilde{A}\right] \notag \\
w_3^y \!&\!=\!&\! -\frac{v^2 \tilde{\tau}}{4\Delta^2} \tilde{A} (\tilde{B}-2), \label{wijsy}
\end{eqnarray}
satisfying $\sum_i w_i^x=1$, and $\sum_i w_i^y=0$. The dashed red lines in Figs.~\ref{Fig4}(d) and (e)
are plotted from Eq.~(\ref{jt}), using the weights in Eqs.~(\ref{wijsx}), and (\ref{wijsy}), whereas the dynamics of the $z$ component, present only for $J \neq 0$, is shown in Fig.~\ref{Fig4}(f).

Turning to the case of strongly coupled TLF with $\Delta \gtrsim v \gg \gamma$, an analysis similar to the one above qualitatively captures the dynamics, although the analytic decay rates in Eqs.~(\ref{G123s}) are accurate only to third order in $v/\Delta$. To illustrate this regime we consider $\delta h=J=0.01\mu$eV, and take $J=\delta h$ or $J=0$. The asymptotic behavior at the short and long $\tau$ limits remains as before, but the crossover from the $\tau^4$ short-time scaling to the constant long-time behavior of $\Gamma_3^s$ occurs at a much larger $\tau$, as seen in Fig.~\ref{Fig5}(a). Since the qubit dynamics are governed by a single decay rate: $\Gamma_1^s=\Gamma_2^s$, which is independent of $\Delta$, the resulting qubit decay depicted by the solid blue line in Fig.~\ref{Fig5}(c) is largely unchanged. In contrast, for $J=0$, the two contributing decay rates scale like $1/\Delta^2$ and are almost two-orders-of-magnitude larger than the rates plotted in Fig.~\ref{Fig4}(b). As a result, a dramatic decrease in coherence time is shown by the dashed red line in Fig.~\ref{Fig5}(c), as compared with the $\Delta \gg v \gg \gamma$ case. Finally, the dissipative dynamics in the $y$ and $z$ components are much more pronounced for the $\Delta \gtrsim v$ case, as seen in Fig.~\ref{Fig5}(d).

\subsection{CPMG}
\label{secCPMG}

Here we extend the above analysis to treat the CPMG protocol, where $\tau_j=t/N$ for $2\leq j \leq N$ and $\tau_1=\tau_{N+1}=t/2N$. The evolution of the qubit state under $N$-pulse CPMG sequence can be written as:
\begin{equation}
{\bf k} (t)= \left\{
\begin{array}{ll} T_{1/2} T^{N-1} T_{1/2}, {\bf k}(0), & N \in {\rm odd} \\[0.2 cm]
 L T_{1/2} T^{N-1} T_{1/2} {\bf k}(0), & N \in {\rm even}, \label{oddeven}
\end{array} \right.
\end{equation}
where $T_{1/2}$ is defined by Eq.~(\ref{Tanal}) with $\tau \rightarrow \tau/2$. Focusing on the $x$ component of the Bloch vector, we consider first a weakly coupled TLF ($v \ll \gamma$), in the case of $J=0$. Here the $z$ component is again decoupled from $x$ and $y$, and we can consider only the $3 \times 3$ block in $T$ corresponding to $(x,y,\delta z)$. Furthermore, since there is no weight for the fast $\sim 2 \gamma$ decay rate, predominantly associated with $\delta z$, in the weak coupling case, we can work within the $2\times 2$ subspace of $(x,y)$.

The similarity transformation that diagonalizes our subset of $T$ is:
\[
S=\left( \begin{array}{cc} \sin \frac{\tilde{\tau}}{2} & -\cos \frac{\tilde{\tau}}{2} \\[0.1 cm]
\cos \frac{\tilde{\tau}}{2} & \sin \frac{\tilde{\tau}}{2} \end{array} \right),
\]
and we can write Eqs.~(\ref{oddeven}) for the $(x,y)$ components as:
\begin{equation}
\left[ \begin{array}{c} x(t) \\ y(t) \end{array} \right] = \tilde{L} S \left( \begin{array}{cc} \chi_1^{N-1} & 0 \\
0 & \chi_2^{N-1} \end{array} \right) S^\dagger  \left[ \begin{array}{c} 1 \\ 0 \end{array} \right],
\end{equation}
where $\chi_i$ are the two eigenvalues of $T$ that are associated with the decay rates in Eqs.~(\ref{G12w}) [recall Eq.~(\ref{eta})], and $\tilde{L}={\rm diag}(1,\pm 1)$, with the upper (lower) sign corresponding to odd (even) number of pulses. Note that $x(t)$ is unaffected by $\tilde{L}$, and we obtain an 'even-odd effect' only for the $y$ component. We find that $x(t)$ decays with a single rate, $\Gamma_2^w$, given in Eqs.~(\ref{G12w}). The well known superior performance of the CPMG protocol over PDD can thus be explained as follows. For $\gamma \ll \Delta$, Eqs.~(\ref{G12w}) can be approximated by $\Gamma_1^w \approx \frac{\gamma v^2}{\Delta^2} (1- {\rm sinc} \tilde{\tau})$, and $\Gamma_2^w \approx \frac{\gamma v^2}{\Delta^2} (1+ {\rm sinc} \tilde{\tau} -2 {\rm sinc}^2 \tilde{\tau}/2)$. An efficient noise suppression is thus achieved when the time interval between pulses satisfies $\tilde{\tau} \ll 1$. At these short time intervals, $\Gamma_1^w \sim \gamma v^2 \tau^2 \gg \Gamma_2^w \sim \gamma v^2 \Delta^2 \tau^4$. Since only $\Gamma_2^w$ is present in the CPMG pulse sequence, it is more effective than PDD at the $\varphi=\pi/2$ point.

Similar calculation can be carried for a general working position. In the general case, we need to include the $z$ component that is now coupled to $x$ and $y$. We find again that the $x$ component decays with a single rate given by:
\begin{equation}
\tilde{\Gamma}^w=\frac{\gamma v^2}{\Delta^2+4 \gamma^2} \left[ (1+A-B_2)\sin^2 \varphi+C\cos^2 \varphi \right], \label{Gcp}
\end{equation}
where $A,B_2$, and $C$ were given in Eqs.~(\ref{A-F}). The $\cos^2 \varphi$ term in Eq.~(\ref{Gcp}) dominates at short $\tau$, and the advantage of CPMG over PDD is largely eliminated at $\varphi \neq \pi/2$.

Turning to the case of a strongly coupled TLF ($v \gg \gamma$), we include the small weight of the fast decaying $(\delta x, \delta y, \delta z)$ components. Considering only the $\varphi=\pi/2$ case, we perform a calculation similar to the above within the subspace $(x,y,\delta z)$, with $\Gamma^s_{1,2}$ given in Eqs.~(\ref{G13}) and the additional small weighted rate of $2\gamma$.  As in the weak coupling case, we find that the weight of $\Gamma_1^s$ is vanished and the resulting qubit decay reads:
\begin{equation}
x(t)=\left(1-\frac{v^2}{\Delta^2} \tilde{A}^2\right)e^{-\Gamma_2^s t}+\frac{v^2}{\Delta^2} \tilde{A}^2 e^{-2\gamma t}, \label{Gcps}
\end{equation}
where $\tilde{A}$ was given in Eqs.~(\ref{A-C}). While the weight of the second term is small, its fast decay rate induces a sizable effect on the qubit decay.

\section{TLF-induced decoherence under general pulse sequence: Results}
\label{results}

In this section we present qubit coherence times due to its coupling to a single TLF. With limited knowledge of the TLF characteristics, we wish to map a wide range of TLF switching rates and coupling strengths, at various qubit bias points. For brevity we focus only on the qubit decay time, $T_2$, which we define as the time it takes the $x$ component of the Bloch vector of a qubit initially prepared along the $x$ axis to reach 50\% of its initial value. We remark that it is often desirable to study the initial qubit dynamics, which may be different for two pulse protocols that have similar $T_2$. In addition, $T_2$ does not capture the dissipative dynamics that leads to nonzero $y$ and $z$ components (see Figs.~\ref{Fig3}-\ref{Fig5}). Nevertheless, the results presented below give good indication for the qubit dynamics in the various regimes.

\begin{figure}[tb]
\epsfxsize=0.675\columnwidth
\vspace*{-0.0 cm}
\centerline{\epsffile{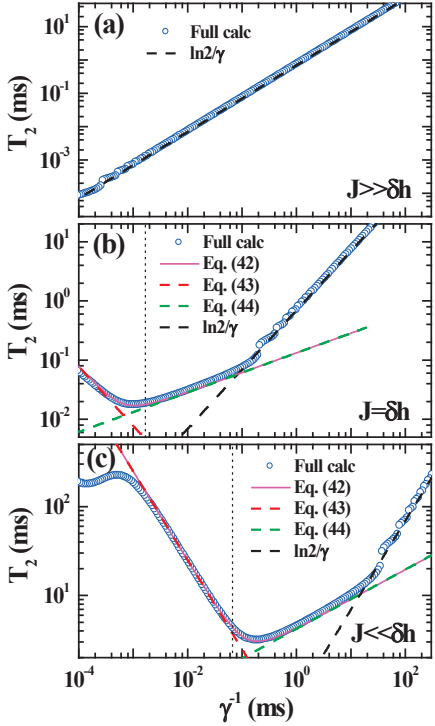}}
\vspace*{-0.2 cm}
\caption{(color online) Qubit decay times vs. TLF average switching times under a 10-pulse CPMG sequence, for a qubit-TLF distance $R=300$ nm. (a) Zero detuning ($J=4.15 \mu$eV, $v=1.4 \mu$eV); (b) moderate negative detuning ($J=0.125\mu$eV, $v=2.5$ neV); and (c) large negative detuning ($J=0.02\mu$eV, $v=62$ peV). In all plots $\delta h=0.125 \mu$eV. The black, green, and red dashed lines correspond to very strong, strong, and weak coupling regimes, respectively. The vertical dotted lines in panels (b) and (c) mark $v=\gamma$ (in panel (a) it is not within range).}
\label{Fig6}
\end{figure}

As a benchmark case, we consider a 10-pulse CPMG sequence. Fig.~\ref{Fig6} shows $T_2$ times vs. TLF average switching times for $R=300$ nm, at three bias points. We identify three coupling strength regimes at which very different dynamics occur, as explained below: (i) very strong coupling, (ii) strong coupling, and (iii) weak coupling, marked by black, green, and red dashed lines, respectively. Note that by weak and strong we refer only to the ratio $v/\gamma$. In Fig.~\ref{Fig6}(a) we consider zero detuning, where $J =4.15 \mu$eV $\gg \delta h$ ($\varphi \approx 0$), $v=1.4 \mu$eV, and the qubit is most susceptible to charge noise. Here we are in the very strong coupling regime for the entire range of TLF switching rates. Since $v/\Delta$ is not small, it is more straightforward to consult the results for pure dephasing given in Appendix A. In this quasi static regime, for our case of 10 pulses, the $\tau$ values relevant for the qubit $T_2$ times satisfy $\tau \sim 1/\gamma \gg 1/v$ for almost the entire range. The large $\tau$ limit of Eq.~(\ref{Gpms}) results in $\Gamma^s_\pm \approx \gamma$ for both rates, thus the qubit approximately decays with a single rate, $\gamma$, and $T_2=\ln 2/\gamma$.

Turning to the case of large negative detuning, where $J=0.02 \mu{\rm eV}\ll \delta h$ ($\varphi \lesssim \pi/2$) and $v=62$ peV, Fig.~\ref{Fig6}(c) includes all three coupling strength regimes. Focusing first on the weak coupling (left side of the vertical dotted line), the qubit decays with a single rate given by Eq.~(\ref{Gcp}). For $\Delta \gg \gamma$, satisfied here, the $\cos^2 \varphi$ term dominates, as long as we are not too close to the $\varphi =\pi/2$ point [see Eqs.~(\ref{A-F})], and the decay rate reads:
\begin{equation}
\Gamma^w =\frac{v^2}{2 \gamma} \left( 1-\frac{\tanh \gamma \tau}{\gamma \tau} \right) \cos^2 \varphi. \label{GW}
\end{equation}
The long- and short-time asymptotes of Eq.~(\ref{GW}) yield the following dephasing times:
\begin{eqnarray}
T_2^{\rm lt} \!&\!=\!&\! \frac{2 \gamma \ln 2}{v^2 \cos^2 \varphi}, \hspace{1.6 cm} \gamma \tau \gg 1,  \label{lt} \\
T_2^{\rm st} \!&\!=\!&\! \left( \!\frac{6 N^2 \ln 2}{\gamma v^2 \cos^2 \varphi} \right)^{1/3}\!\! , \hspace{0.5 cm}  \gamma \tau \ll 1 \label{st}
\end{eqnarray}
where $N$ is the number of pulses. Observing the solid magenta line in Fig.~\ref{Fig6}(c) that depicts Eq.~(\ref{GW}), and the dashed red and green lines, corresponding to its asymptotes, we notice that Eq.~(\ref{GW}) works well into the strong coupling regime. This can be explained by taking the large $\tau$ limit of Eqs.~(\ref{G123s}), where we have found a single decay rate: $\Gamma^s=\gamma v^2 \tau^2 \cos^2 \varphi/6$, which is identical to the large $\tau$ limit of Eq.~(\ref{GW}). For very weak coupling [left end of Fig.~\ref{Fig6}(c)], the approximate result of Eq.~(\ref{GW}) breaks down since $\Delta \gg \gamma$ no longer holds. We note that although $J \ll \delta h$, the dynamics shown here is very different from the case of $J=0$. In the latter case, $T_2$ times increase substantially, due to much smaller rates given by Eq.~(\ref{Gcp}). To realize this limit one needs to work at a very large negative bias, or to utilize inter-qubit couplings that can turn off the exchange completely. Fig.~\ref{Fig6}(b) shows decay times for moderate negative bias where $J=\delta h$, exhibiting the complex dynamics discussed above.

\begin{figure}[tb]
\epsfxsize=0.7\columnwidth
\vspace*{-0.0 cm}
\centerline{\epsffile{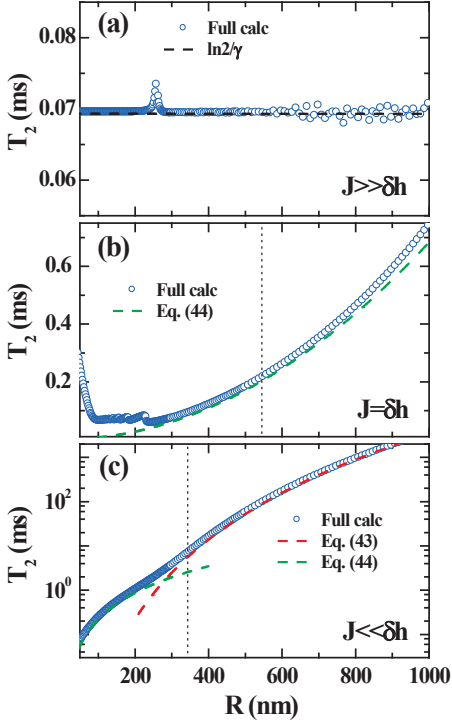}}
\vspace*{-0.14 cm}
\caption{(color online) Qubit decay times vs. qubit-TLF distance under a 10-pulse CPMG sequence, for a TLF average switching time of 0.1 ms. (a) Zero detuning ($J=4.15 \mu$eV); (b) moderate negative detuning ($J=0.125\mu$eV); and (c) large negative detuning ($J=0.02\mu$eV). In all plots $\delta h=0.125 \mu$eV. The black, green, and red dashed lines correspond to very strong, strong, and weak coupling regimes, respectively. The vertical dotted lines in panels (b) and (c) mark $v=\gamma$.}
\label{Fig7}
\end{figure}
In Fig.~\ref{Fig7} we show $T_2$ times vs. qubit-TLF distance, for a TLF average switching time of 0.1 ms, at the same three bias points. At zero detuning we are again at the very strong coupling regime throughout the considered range, resulting in a constant $T_2=\ln 2/\gamma =69.3 \mu$s, as shown in Fig.~\ref{Fig7}(a). At large negative detuning, Fig.~\ref{Fig7}(c) exhibits the weak and strong coupling regimes captured by Eqs.~(\ref{lt}), and (\ref{st}). Here, the relevant $\tau$ values scale as $1/v$. To the left of the vertical dotted line, marking $v=\gamma$, we have strong coupling, thus $\gamma \tau \ll 1$, and Eq.~(\ref{st}) applies. Since the dominant term in the qubit-TLF coupling scales as $R^{-3}$ [see Fig.~\ref{Fig1}(c) and discussion therein], $T_2$ scales  like $R^2$ in this regime. To the right of the $v=\gamma$ line we have $\gamma \tau \gg 1$, and Eq.~(\ref{lt}) applies, thus $T_2$ scales as $R^6$. The $T_2$ times in the intermediate bias position shown in Fig.~\ref{Fig7}(b), are understood by the same arguments all the way down to $R \approx 250$ nm, below which, the very strong coupling, quasi-static regime applies, and $T_2$ is roughly constant. The increase in $T_2$ times below $R=100$ nm is not explained within our analytic results, which do not apply to this $v \gg \Delta$ regime.

\begin{figure}[tb]
\epsfxsize=0.7\columnwidth
\vspace*{-0.0 cm}
\centerline{\epsffile{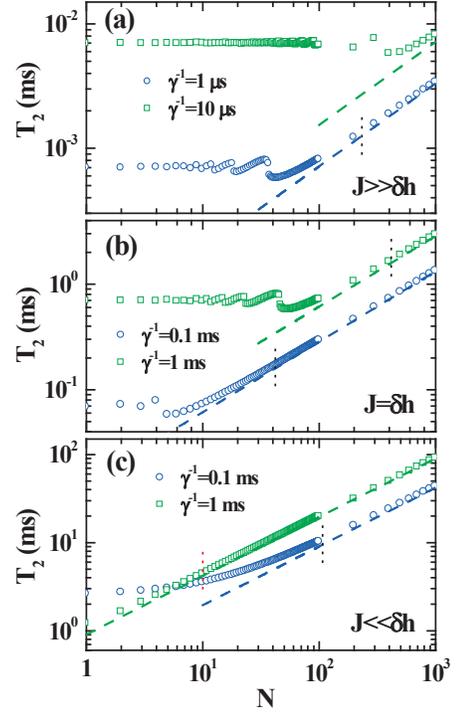}}
\vspace*{-0.2 cm}
\caption{(color online) Qubit decay times vs. number of CPMG pulses for $R=300$ nm. (a) Zero detuning ($J=4.15 \mu$eV); (b) moderate negative detuning ($J=0.125\mu$eV); and (c) large negative detuning ($J=0.02\mu$eV). In all plots $\delta h=0.125 \mu$eV. The dashed lines correspond to Eq.~(\ref{st}). The vertical dotted black (red) lines mark the crossover from very strong (weak) to strong coupling regime.}
\label{Fig8}
\end{figure}
The discussion above suggests that prolonging the qubit coherence time by reducing the time interval between pulses (or equivalently increasing the number of pulses) is effective only when Eq.~(\ref{st}) holds. Within this regime, marked by the overlap of the dashed green lines with the full calculation in Figs.~\ref{Fig6}, and \ref{Fig7}, decay time scales with the number of pulses as $N^{2/3}$. This power law was predicted for pure dephasing due to random telegraph noise in Ref.~\onlinecite{Cywinski2}. There it was explained by observing that as $N$ increases, the non-Gaussian noise attributes are suppressed by the DD pulses. The $N^{2/3}$ scaling was then found by considering Gaussian noise with spectral density having a soft cutoff (e.g. $\omega^{-2}$).\cite{Cywinski2} Furthermore, the same power law was observed in experiments on spin qubits in nitrogen vacancy centers in diamond, where both CPMG and double-axis (XY) DD pulse sequences were applied.\cite{Lange} In this system, the dephasing of the central spin is induced by its coupling to a bath of spins with dipolar intra-bath coupling. By identifying the noise as a Gaussian and Markovian Ornstein-Uhlenbeck process, the authors were able to explain the $N^{2/3}$ scaling using the arguments given in Ref.~\onlinecite{Cywinski2}.

We find estimates for the number of pulses required to crossover into the large-$N$ regime from either side. At the very strong coupling regime (dashed black lines), $T_2 = \ln 2/\gamma$, therefore to avoid the large $\tau$ limit of Eq.~(\ref{Gpms}), we require $v \tau \approx (v/\gamma) \ln 2/N \lesssim 1$, or $N \gtrsim (v/\gamma) \ln 2$. At the weak to strong coupling crossover we have $T_2 = \ln 2/\Gamma^{\rm st}$, given by Eq.~(\ref{st}), and to satisfy the condition $\gamma \tau \lesssim 1$, we need $N \gtrsim 6 \ln 2(\gamma/v \cos \varphi)^2$. To summarize, the minimum number of pulses needed to enter the regime at which $T_2$ scales as $N^{2/3}$ is estimated by:
\begin{equation}
N \gtrsim {\rm max} \left( \frac{v\ln 2}{\gamma}, \frac{6 \ln 2}{\cos^2 \varphi} \frac{\gamma^2}{v^2} \right). \label{Nmax}
\end{equation}

These considerations are demonstrated in Fig.~\ref{Fig8}, where we plot $T_2$ times vs. the number of CPMG pulses for a qubit-TLF distance of $R=300$ nm, at the above three bias points. The dashed lines correspond to the power law: $T_2 \propto N^{2/3}$. The vertical dotted lines mark the minimum number of pulses necessary to enter the above regime, where black and red lines correspond to the left and right conditions in Eq.~(\ref{Nmax}), respectively. While the $N^{2/3}$ power law is in effect only for $N$ satisfying Eq.~(\ref{Nmax}), considerable increase in coherence time is still obtained for smaller number of pulses, to the left of the vertical dotted lines. It is interesting to note that the 2/3 power law is close to the value 0.72, recently found for even number of CPMG pulses in a setting that was likely dominated by nuclear-induced dephasing.\cite{Medford} Finally, we comment that we have assumed that our control pulses are ideal, in that they are zero-width $\pi$ pulses applied exactly along the $y$ axis. In reality, control pulses have errors in both their rotation angle and axis, which may introduce more noise than they can effectively remove.\cite{BerFao,Wang} This is particularly relevant for large-$N$ sequences, thus the estimates in Fig.~\ref{Fig8} should not be taken too seriously above $N \approx 50$.

\begin{figure}[tb]
\epsfxsize=0.95\columnwidth
\vspace*{-0.05 cm}
\centerline{\epsffile{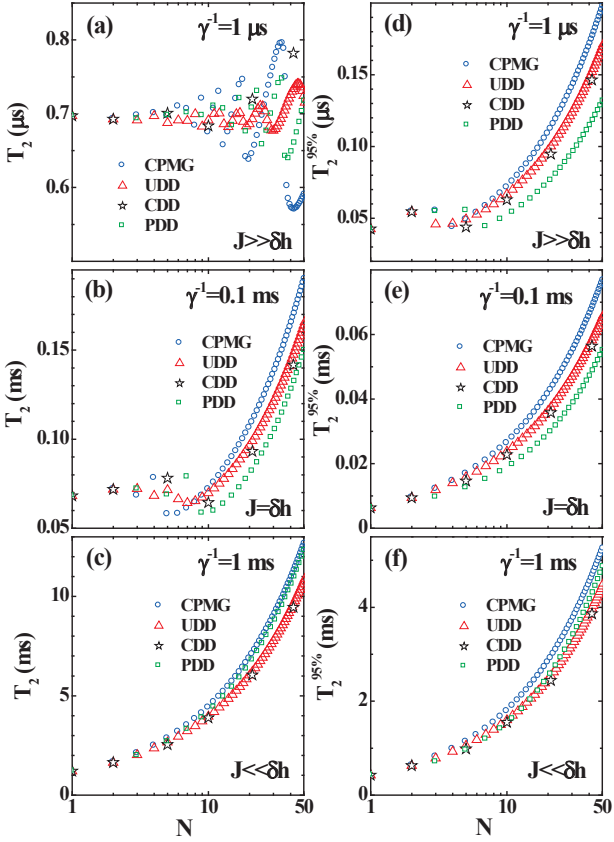}}
\vspace*{-0.1 cm}
\caption{(color online) Qubit decay times vs. number of pulses for $R=300$ nm under various protocols. (a) Zero detuning ($J=4.15 \mu$eV), TLF average switching time of $1 \mu $s; (b) moderate negative detuning ($J=0.125\mu$eV), TLF average switching time of $0.1$ ms; and (c) large negative detuning ($J=0.02\mu$eV), TLF average switching time of 1 ms. Panels (d), (e), and (f) show decay times to 95\% of initial value, for same parameters as in panels (a), (b), and (c), respectively. In all plots $\delta h=0.125 \mu$eV. ${\rm CDD}_2$ and 2-pulse UDD are the same as 2-pulse CPMG, and PDD has only odd number of pulses.}
\label{Fig9}
\end{figure}
Next we compare $T_2$ times under several widely used DD protocols. The concatenated DD sequence (CDD) is defined recursively,\cite{Kaveh} with the $l$th order of concatenation given by:
\begin{equation}
{\rm CCC}_l (t)={\rm CDD}_{l-1} (t/2) - \pi - {\rm CDD}_{l-1} (t/2),
\end{equation}
where ${\rm CDD}_0$ is free induction, and the middle $\pi$ pulse is included only for odd $l$.
The number of pulses in the $l$th level concatenation is found as:
\begin{equation}
N_l=\left\{ \begin{array}{ll} \frac{1}{3} \left(2^{l+1}-1\right), & l\in {\rm odd} \\[0.2 cm]
 \frac{1}{3} \left(2^{l+1}-2\right), & l\in {\rm even}. \end{array} \right.
\end{equation}
The Uhrig DD sequence (UDD) is defined by pulse times:
\begin{equation}
t_i=t\sin^2 \left[\frac{i\pi}{2(N+1)}\right], \hspace{0.5 cm} 1\leq i \leq N
\end{equation}
where the time interval between the $(i-1)$th and $i$th pulses is $\tau_i=t_i -t_{i-1}$ ($t_0=0$, $t_{N+1}=t$).\cite{Uhrig} We have not attempted to obtain analytical results for the decay rates of pulse sequences with unequal time intervals between pulses. In Fig.~\ref{Fig9} we compare the performance of PDD, CPMG, CDD, and UDD sequences at the above three bias points. At zero detuning, Fig.~\ref{Fig9}(a) shows no advantage in increasing the number of control pulses up to $N=50$, for any of the protocols, even though we are considering a relatively fast TLF ($\gamma^{-1}=1 \mu s$). At this strong coupling limit, the qubit state decay is characterized by plateaus (see, e.g., Fig.~\ref{Fig10}) that result in the observed oscillations in the dependence of $T_2$ on $N$. A better performance measure may be decay time to $95\%$ of initial value, plotted in Fig.~\ref{Fig9}(d), where coherence times are consistently improved with the number of pulses. Decay times at moderate- and large-negative bias points are shown in Figs.~\ref{Fig9}(b) and (c), respectively, with their respective 5\% drop times given in Figs.~\ref{Fig9}(e) and (f). Throughout the wide TLF parameter range considered, and at all qubit bias points, the CPMG protocol exhibits superior performance in mitigating charge noise. These results agree and complement previous observations of the superiority of CPMG over UDD in fighting noise with power-law high-frequency tail.\cite{Cywinski2,Pasini}  We also note that all pulse protocols roughly follow the $N^{2/3}$ power law over a wide range of TLF parameters and qubit working positions.

\section{Conclusion}

In this work we have studied the dynamics of a singlet-triplet spin qubit afflicted by a charge fluctuator, under dynamical decoupling control pulses. We have presented a theory that predicts rich dynamics governed by the qubit-TLF coupling strength and the TLF switching rates, as well as by the qubit working position. For a relatively small fixed magnetic field gradient, the exchange interaction at zero detuning results in pure dephasing that is predominantly dependant on the TLF parameters. In contrast, at negative bias points the qubit dynamics becomes dissipative and the effectiveness of the DD pulses depends on both TLF and qubit characteristics.

We have demonstrated that $\pi_y$ pulses are preferable over $\pi_x$ pulses in several scenarios, and moreover we expect that $\pi_y$ pulses will be superior in eliminating nuclear fluctuations. Finding analytical formulas for the qubit decay rates in the limits of weak and strong coupling for PDD and CPMG sequences, enabled us to explain our results for qubit coherence times as functions of TLF distance and switching rate at various bias points. Over a large range of system parameters, coherence times follow a  power law $T_2 \propto N^{2/3}$, where $N$ is the number of pulses. In addition, comparing the performance of several pulse protocols, we found that CPMG is the most effective protocol to eliminate charge noise. While we have presented specific results for two-spin qubits in gate-defined GaAs QDs, the formulation presented in this paper should be relevant to a wide variety of systems afflicted by charge fluctuators. A natural extension of our work will include pulse-error analysis.\cite{BerFao,Wang} In particular, the tolerance of various pulse sequences to accumulation of systematic and random pulse errors should be evaluated in the context of charge noise.

As current experiments move to more complicated QD structures, such as two coupled double QDs,\cite{Weperen,Shulman} and three-spin qubits,\cite{Laird,Gaudreau} we expect that charge noise will play an increasing role in the system decoherence. There are two main aspects needed to be considered when analyzing charge noise in larger devices. First, as the sample size increases, larger number of active charge traps can couple to the qubit(s). It is well known that a collection of weakly-coupled TLFs with a broad distribution of switching rates can lead to $1/f$ noise spectrum. On the other hand, qubit decoherence due to small mesoscopic ensembles can be dominated by the strongest fluctuator(s), and the generated noise is non-Gaussian with large variability between samples. The system dynamics can thus be quite different when crossing over from small to larger devices. A second point is that coupled qubits may provide access to working points that are unavailable for a single qubit. As the interdot bias controls simultaneously the exchange interaction and TLF coupling strength, the results of section \ref{results} were restricted to the available subset of system parameters. With two or more qubits, inter-qubit couplings provide an additional handle over $J$, which is unrelated to the TLF coupling. In particular, optimal bias points at which $J=0$, become available, where a very different dynamics is predicted.

Finally, experimental measurement of the qubit coherence time under DD sequences can provide insight into the spectral characteristics of the dominant noise processes.\cite{Biercuk,Yuge} The distinctive dynamics found for weak and strong fluctuators under DD, can therefore facilitate the characterization of charge fluctuators in solid-state devices. Moreover, by comparing our calculated decay rates for a general working point, where dissipative dynamics occur, with the results of a theory based on a Gaussian and Markovian noise, one can find both dephasing and relaxation functions, and determine the validity range of a Gaussian noise theory for qubit-TLF coupling under dynamical decoupling.

\section*{ACKNOWLEDGEMENTS}

The author thanks Joakim Bergli for helpful discussions and acknowledges funding from Research Corporation.

\section*{APPENDIX A: PURE DEPHASING}
\renewcommand{\theequation}{A\arabic{equation}}
\setcounter{equation}{0}

For pure dephasing ($\varphi =0$), Eq.~(\ref{Bpm}) reads ${\bf B}_\pm =(0,0,J\mp v)$. This case is realized when $\delta h=0$ and has been studied extensively. We outline its solution here to provide context to the general case studied in the main text. The controlled $z$ rotation generated by $J$ is eliminated in all cases except for free induction, thus we disregard $J$ with the understanding that a qubit initially prepared along the $x$ axis acquires only a random phase. Since we need to consider only the $x$ component dynamics, it is more straightforward to evaluate the signal decay by:

\begin{equation}
\chi (t) = \int d\phi p(\phi,t) e^{i \phi}.
\end{equation}
Dividing the probability distribution to $p_\pm (\phi,t)$ to accumulate phase $\phi$ while the TLF is in the up or down state, we can write a set of coupled equations for $\chi_\pm$, analogous to Eqs.~(\ref{ppm}):
\begin{equation}
\dot{\chi}_\pm= -\gamma_\pm \chi_\pm+\gamma_\mp \chi_\mp \mp iv \chi_\pm.
\end{equation}
These can be converted into equations for $\chi=\chi_+ +\chi_-$, and $\delta \chi=\chi_+ -\chi_-$:
\begin{eqnarray}
\dot{\chi}\!&\!=\!&\! -iv \delta \chi \notag \\
\dot{\delta \chi} \!&\!=\!&\! -\delta \gamma \chi -2 \gamma \delta \chi -iv \chi,
\end{eqnarray}
which, in turn, are transformed into a second order equation for $\chi$:
\begin{equation}
\ddot{\chi}+2 \gamma \dot{\chi}+v^2\chi-iv \delta \gamma \chi =0, \label{chi}
\end{equation}
with initial conditions $\chi (t=0) =1$, and $\dot{\chi} (t=0)=iv \tanh(\Delta E_t/2k_BT)$.
The general solution to Eq.~(\ref{chi}) is\cite{Laikhtman,Paladino,BerFao}
\begin{equation}
\chi (t)=a_+ e^{-\gamma(1-\mu)t}+a_- e^{-\gamma(1+\mu)t}, \label{chisol}
\end{equation}
where
\begin{eqnarray}
\mu\!&\!=\!&\! \sqrt{1-\left(\frac{v}{\gamma}\right)^2+\frac{2iv}{\gamma} \tanh \left(\frac{\Delta E_T}{2k_B T} \right)} \notag \\
a_\pm\!&\!=\!&\! \frac{(\mu_R \pm 1)(1 \pm i\mu_I)}{2 \mu}, \label{muab}
\end{eqnarray}
and $\mu_R, \mu_I$ are the real and imaginary parts of $\mu$. The same result is obtained from the more general formulation presented in section \ref{seciii}. For pure dephasing with qubit initially along the $x$ axis, $z_\pm (t)=0$, and the remaining four equations in Eqs.~(\ref{xyz}) decouple into two sets $(x, \delta y)$, and $(\delta x, y)$, with a solution given by appropriately rotating Eqs.~(\ref{chisol})-(\ref{muab}).

In order to consider the effects of a sequence of $\pi$ pulses we first note that in the case of pure dephasing, $\pi_x$ and $\pi_y$ pulses flip the qubit state in the same way, since the qubit evolution involves only the $\sigma_z$ operator ($\delta h=0$). Repeated applications of $\pi$ pulses alternately change the sign of the last term in Eq.~(\ref{chi}), resulting in a solution similar to Eq.~(\ref{chisol}) with $\mu \rightarrow \mu^*$. A solution for the qubit signal decay after the application of a general sequence of $\pi$ pulses can be found by stitching the solutions between pulses. This is done using a transfer matrix approach analogous to Eqs.~(\ref{ft})-(\ref{T}). For a sequence of $N$ pulses we find:
\begin{equation}
\left( \begin{array}{c} a_+^{j+1} \\ a_-^{j+1} \end{array} \right) = \Lambda_{\rm o} (\tau_j)\left( \begin{array}{c} a_+^{j} \\ a_-^{j} \end{array} \right), \hspace{0.5 cm} j \in {\rm odd} \label{odd}
\end{equation}
and
\begin{equation}
\left( \begin{array}{c} a_+^{j+1} \\ a_-^{j+1} \end{array} \right) = \Lambda_{\rm e} (\tau_j) \left( \begin{array}{c} a_+^{j} \\ a_-^{j} \end{array} \right), \hspace{0.5 cm} j \in {\rm even}
\end{equation}
where
\begin{equation}
\Lambda_{\rm o}(\tau_j)\!=\!\frac{e^{-\gamma (\mu +1) \tau_j}}{\mu^*} \!\!\left(\!\! \begin{array}{cc} e^{2 \gamma \mu \tau_j} (1-i \mu_I) \!&\! 1+\mu_R \\  e^{2 \gamma \mu \tau_j} (\mu_R -1) \!&\! -(1+i\mu_I) \end{array} \!\! \right)\!\!,
\end{equation}
$\tau_j$ is the time interval between pulses $j$ and $j+1$, $\Lambda_{\rm e}=\Lambda_{\rm o}^*$, and $a_\pm^1$ are given by Eq.~(\ref{muab}). The final solution for the qubit signal decay after $N$ control pulses reads
\begin{equation}
\chi_{dd} (t)=a_+^{N+1} e^{-\gamma(1-\mu)t}+a_-^{N+1} e^{-\gamma(1+\mu)t}, \label{chidd}
\end{equation}
for even $N$, and a similar solution with $\mu \rightarrow \mu^*$ for odd $N$. As an example we retrieve below the known result for SE, as well as the signal decay in the case of two-pulse CPMG
\begin{eqnarray}
\chi_{\rm SE} (t)\!&\!=\!&\! \frac{e^{-\gamma t}}{2|\mu|^2} \left[ (\mu_I^2+1) \sum_\pm (1\pm \mu_R) e^{\pm \gamma \mu_R t} \right. \nonumber \\
\!&\!+\!&\! \left. (\mu_R^2-1) \sum_\pm (1 \pm i\mu_I) e^{\pm i\gamma \mu_I t} \right], \label{chise}
\end{eqnarray}
\begin{eqnarray}
\chi_{\rm CP_2} (t)\!&\!=\!&\! \frac{e^{-\gamma t}}{2\mu|\mu|^2} \left\{ 4(\mu_R^2-1)(\mu_I^2+1) \sinh \left(\frac{\gamma \mu^*t}{2}\right) \right. \nonumber \\
\!&\!+\!&\! \left. \sum_\pm (\mu_R \pm 1)(1\pm i \mu_I) \left[ (\mu_R^2-1)e^{\pm i\gamma \mu_I t} \right. \right.\notag \\
\!&\!+\!&\! \left. \left. (1+\mu_I^2) e^{\pm \gamma \mu_R t} \right] \right\}. \label{chicp2}
\end{eqnarray}

For $N$ pulse PDD or CPMG sequences, analytical results are presented in section \ref{anal}. For PDD with symmetric TLF switching, it can be shown that the qubit dephasing is given by two exponentials, similarly to Eq.~(\ref{chidd}). These exponentials are the eigenvalues of the matrix $\Lambda_{\rm e} (\tau) \Lambda_{\rm o} (\tau)$, where $\tau= t/(N+1)$. They are found as:\cite{BerFao}
\begin{equation}
\chi_\pm =\frac{e^{-\gamma \tau}}{\mu} \left( \sinh \gamma \mu \tau \pm \sqrt{ \cosh^2 \gamma \mu \tau -\frac{v^2}{\gamma^2}} \right). \label{chipm}
\end{equation}
In the limit of weak coupling, $v \ll \gamma$, the $\chi_+$ eigenvalue dominates, and we obtain a single decay rate by using Eq.~(\ref{eta}):
\begin{equation}
\Gamma^w = \frac{v^2}{2\gamma} \left( 1-\frac{\tanh \gamma \tau}{ \gamma \tau} \right).
\end{equation}
This result was found in the main text, by taking $\varphi=0$ in Eqs.~(\ref{G123w}). and (\ref{A-F}).

In the limit of strong coupling, $v \gg \gamma$, Eq.~(\ref{chipm}) results in two oscillating rates:
\begin{equation}
\Gamma_\pm^s = \gamma \left( 1 \pm {\rm sinc} v \tau \right), \label{Gpms}
\end{equation}
which contribute to the qubit dephasing with corresponding weights:
\begin{equation}
w_\pm =\frac{1}{2} (1\pm \cos v \tau).
\end{equation}
The short $\tau$ limit of Eq.~(\ref{Gpms}) can be retrieved from Eqs.~(\ref{G123s}) and (\ref{A-C}) in the main text, where we presented the strong coupling results for the case of $v \ll \Delta$.

\begin{figure}[t]
\epsfxsize=0.7\columnwidth
\vspace*{0 cm}
\centerline{\epsffile{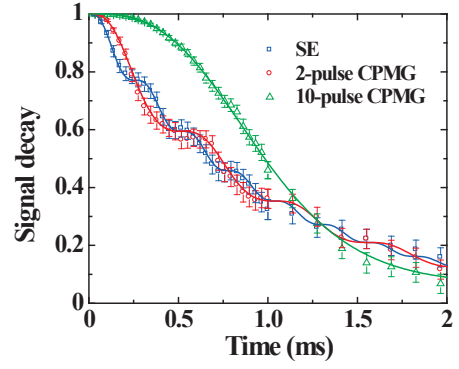}}
\vspace*{-0.2 cm}
\caption{(color online) Signal decay for $\varphi =0$ under SE (blue line, circles), 2-pulse CPMG (red line, triangles), and 10-pulse CPMG (green line, squares). Solid lines show analytic results and symbols depict averages and standard deviations of 1000 numerical realizations of the TLF random switching process. In all plots $\delta h=0$, $v=0.1$ neV, average TLF switching time is $\gamma =1$ ms, and TLF level splitting is $\Delta E_T=0.45 \mu$eV.}
\label{Fig10}
\end{figure}
Fig.~\ref{Fig10} shows signal decay for several DD sequences, calculated for a TLF with an average switching time of 1 ms, and coupling strength of $v=0.1$ neV. The solid blue and red lines correspond to SE and 2-pulse CPMG given by Eqs.~(\ref{chise}) and (\ref{chicp2}), respectively. The green line, depicting 10-pulse CPMG, was generated using the transfer-matrix method, Eqs.~(\ref{odd})-(\ref{chidd}). To verify our analytical results, we performed a numerical simulation of the TLF random switching, using a Poisson process with time constants $\gamma_\pm$. The symbols in Fig.~\ref{Fig10} depict an average over 1000 realizations of the random switching, with appropriately weighted initial TLF states. We have carried similar simulations to verify our results for other DD sequences at general working positions, presented in the main text.

\section*{APPENDIX B: Effects of frame rotation on qubit dynamics}
\renewcommand{\theequation}{B\arabic{equation}}
\setcounter{equation}{0}

The analysis presented in the main text was applied to find the dynamics of the qubit state components in the rotated frame. The initial qubit state defined in Eq.~(\ref{k0}) is thus appropriately set to lie along the rotated $x$ axis. Current experimental techniques in QD $S-T_0$ spin qubits have limited accessibility to this state. The purpose of this appendix is to provide a more direct connection to currently investigated $S-T_0$ qubits by including the effects of the frame rotation. In this appendix only the Bloch components are defined in the original frame, thus the $x$ axis refers to the state $(|S\rangle +|T_0 \rangle)/\sqrt{2}$. Rotated quantities stated in the main text are denoted here with a prime. We note that for pure dephasing ($\varphi =0$) the two frames coincide and all our previous results are intact. The qubit dynamics at the anticrossing point, where $J \gg \delta h$ and charge noise dominates, is thus adequately captured in the main text.

Eq.~(\ref{Hfinal}), from which our dynamical equations are derived is defined in a frame rotated by an angle $\varphi$ in the $x-z$ plane. Vectors and matrices are transformed to this frame by:
\begin{eqnarray}
{\bf k}' &=& {\cal O} {\bf k} \\
M' &=& \mathcal{O} M \mathcal{O}^{-1},
\end{eqnarray}
where
\begin{equation}
\mathcal{O}\!=\!\left(\! \begin{array}{cccccc} \cos \varphi \!&\! 0 \!&\! 0 \!&\! 0 \!&\! 0 \!&\! -\sin \varphi \\ 0 \!&\! 1 \!&\! 0 \!&\! 0 \!&\! 0 \!&\! 0 \\ 0 \!&\! 0 \!&\! \cos \varphi \!&\! \sin \varphi \!&\! 0 \!&\! 0 \\ 0 \!&\! 0 \!&\! -\sin \varphi \!&\! \cos \varphi \!&\! 0 \!&\! 0 \\
0 \!&\! 0 \!&\! 0 \!&\! 0 \!&\! 1 \!&\! 0 \\ \sin \varphi \!&\! 0 \!&\! 0 \!&\! 0 \!&\! 0 \!&\! \cos \varphi \end{array} \!\right)\!. \label{O}
\end{equation}

In the following we consider a qubit initially prepared along the $x$ axis: $|\psi (t=0) \rangle =(|S\rangle +|T_0 \rangle)/\sqrt{2}$. In the rotated frame the initial state replacing Eq.(\ref{k0}) reads:
\begin{equation}
{\bf k}'(0)=\left( \cos \varphi,0,-\sin\varphi \frac{\delta \gamma}{2 \gamma},-\cos \varphi \frac{\delta \gamma}{2 \gamma},0,\sin \varphi \right).
\end{equation}
To find the qubit's evolution under $N$-pulse PDD control sequence we rotate Eq.~(\ref{kt}) back to the original frame:
\begin{equation}
{\bf k}(t)= \mathcal{O}^{-1} T^{N+1} \mathcal{O} {\bf k} (0),
\end{equation}
where $T$ is defined in Eq.~(\ref{Tanal}). The decay rates, $\Gamma_i$, found in Sec.~IV from the eigenvalues of $T$, are unaffected by the above rotation, and only their weights in the solution for the Bloch components are modified. In order to find the transformed weights we use the eigenvectors of $T$, as described in Sec.~IVA. 

For the case of weak coupling the weights of the three decay rates, Eqs.~(\ref{G123w}), in the three (non-rotated) Bloch components are found as:
\begin{eqnarray}
w_1^x \!&\!=\!&\! \cos^2 \varphi \sin^2 \frac{\tilde{\tau}}{2} \notag \\
w_{2,3}^x \!&\!=\!&\! \frac{1}{2} \left[ \cos^2 \varphi \cos^2 \frac{\tilde{\tau}}{2} \left(1 \pm \frac{F}{\sqrt{F^2+D^2 \sin^2 2 \varphi}} \right) + \right. \notag \\ && \left. \sin^2 \varphi \left(1 \mp \frac{F}{\sqrt{F^2+D^2 \sin^2 2 \varphi}} \right) \mp \right. \notag \\[-0.05 cm] && \left. \sin^2 2\varphi \cos \frac{\tilde{\tau}}{2} \frac{D}{\sqrt{F^2+D^2 \sin^2 2\varphi}} \right] \notag \\
w_1^y \!&\!=\!&\! \frac{1}{2} \cos \varphi \sin \tilde{\tau} \notag \\
w_{2,3}^y \!&\!=\!&\! -\frac{1}{4} \cos \varphi \sin \tilde{\tau} \left(1 \pm \frac{F}{\sqrt{F^2+D^2 \sin 2 \varphi}} \right) \mp \notag \\ && \frac{1}{2} \sin \varphi \sin \frac{\tilde{\tau}}{2} \frac{D \sin 2\varphi}{\sqrt{F^2+D^2 \sin^2 2\varphi}} \notag \\
w_1^z \!&\!=\!&\! -\frac{1}{2} \sin 2\varphi \sin^2 \frac{\tilde{\tau}}{2} \notag \\
w_{2,3}^z \!&\!=\!&\! \mp \frac{1}{4} \cos \frac{\tilde{\tau}}{2} \frac{D \sin 4 \varphi}{\sqrt{F^2+D^2 \sin 2 \varphi}} + \notag \\ && \frac{1}{4} \sin 2\varphi \left[ \left(1 \mp \frac{F}{\sqrt{F^2+D^2 \sin^2 2 \varphi}} \right) - \right. \notag \\ && \left. \cos^2 \frac{\tilde{\tau}}{2}  \left(1 \pm \frac{F}{\sqrt{F^2+D^2 \sin^2 2 \varphi}} \right) \right], \label{wijwt}
\end{eqnarray}
where $\tilde{\tau}$, $D$, and $F$ are defined in Eqs.~(\ref{A-F}), and $\sum_i w_i^x=1$, and $\sum_i w_i^y=\sum_i w_i^z=0$. Note that the rotation matrix mixes $x$ and $z$ components only when $\varphi \neq 0,\pi/2$, thus dissipative dynamics is minimal when $\delta h \gg J$ (large negative detuning) or $J \gg \delta h$ (at the singlet anticrossing).

For strong coupling there is a nonzero weight for the fast $2 \gamma$ decay rate predominantly associated with the dynamics of $(\delta x, \delta y, \delta z)$,  in addition to the contributions from the three rates given in Eqs.~(\ref{G123s}). For $\varphi=\pi/2$ the transformation in and out of the rotated frame eliminates the weights of the two rates given in Eqs.~(\ref{G13}), and instead the dynamics is governed by the rate:
\begin{equation}
\Gamma'^s_1 (\varphi=\pi/2) =\frac{2 \gamma v^2}{\Delta^2} \left( 1-{\rm sinc}^2 \frac{\tilde{\tau}}{2} \right),
\end{equation}
and an additional non-zero contribution from $\Gamma'^s_2=2\gamma$. The weights of these two rates in the solution for the $x$ component are found as:
\begin{eqnarray}
w_1^x \!&\!=\!&\! 1-\frac{2 v^2}{\Delta^2} \left( \tilde{A}-\frac{\tilde{B}}{4} \right) \notag \\
w_2^x \!&\!=\!&\! \frac{2 v^2}{\Delta^2} \left( \tilde{A}-\frac{\tilde{B}}{4} \right),
\end{eqnarray}
where $\tilde{A}$ and $\tilde{B}$ are defined in Eqs.~(\ref{A-C}). Note that at $\varphi =\pi/2$ there is no dynamics in the $y$ and $z$ axes for our initial state.

\begin{figure}[t]
\epsfxsize=0.9\columnwidth
\vspace*{0.0 cm}
\centerline{\epsffile{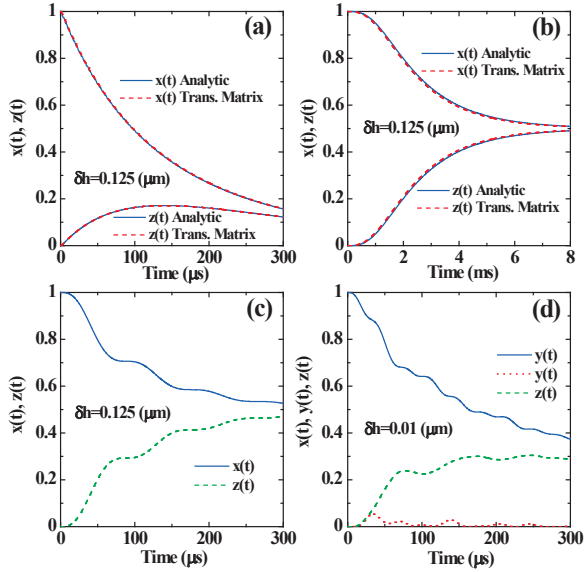}}
\vspace*{-0.1 cm}
\caption{(color online) Non-rotated Bloch vector components under 11-pulse PDD sequence for working point $\delta h =J$ ($\varphi=\pi/4$). (a) Weakly-coupled fast TLF ($1/\gamma=0.1 \mu s$, $v=2.5$ neV), $\Delta \sim \gamma \gg v$; (b) Weakly-coupled slow TLF ($1/\gamma=0.1$ ms, $v=8.4$ peV); $\Delta \gg \gamma \gg v$; (c) Strongly-coupled TLF ($1/\gamma=0.1$ ms, $v=2.5$ neV, $\delta h=0.125 \mu eV$), $\Delta \gg v \gg \gamma$; (d) Strongly-coupled TLF ($1/\gamma=0.1$ ms, $v=2.5$ neV, $\delta h=0.01 \mu eV$), $\Delta \gtrsim v \gg \gamma$.}
\label{Fig11}
\end{figure}

Fig.~\ref{Fig11} shows the time evolution of the Bloch vector components in the non-rotated frame under 11-pulse PDD sequence. Fig.~\ref{Fig11}(a) captures a weakly-coupled fast TLF with parameters as in Figs.~\ref{Fig3}(b) and (c), while Fig.~\ref{Fig11}(b) shows a weakly-coupled slow TLF with parameters as in Fig.~\ref{Fig3}(e). The dashed red lines depict the transfer matrix calculation while the solid blue lines correspond to the analytic solution given by Eqs. (\ref{jt}), (\ref{G123w}), and (\ref{wijwt}). In both cases the most notable difference with respect to the results of Fig.~\ref{Fig3} is the prominent dissipative dynamics associated with the $z$ component. The $y$ component (not shown) is an order-of-magnitude larger than its counterpart in Fig.~\ref{Fig3}, but is still considerably smaller.

Similarly, Figs.~\ref{Fig11}(c), and (d) show qubit dynamics under 11-pulse PDD sequence for a strongly-coupled TLF, with parameters corresponding to Figs.~\ref{Fig4} and \ref{Fig5}, respectively. The mixing between the $x$ and $z$ components, caused by the rotation, is apparent in the strong case as well. The $y$ component shows appreciable dynamics only when $\Delta \gtrsim v$, as seen by the dotted red line in Fig.~\ref{Fig11}(d). For finite $\delta h$ and zero exchange ($\varphi =\pi/2$) the $x$ component in all cases shows comparable (slightly faster) decay to those given in Figs.~\ref{Fig3}(f), \ref{Fig4}(d), and \ref{Fig5}(c). As seen from Eqs.~(\ref{wijwt}) there is no relaxation in this case ($y(t)=z(t)=0$).

\bibliographystyle{amsplain}

\begin{thebibliography}{99}
\bibitem{angular} A more accurate angular averaging would be to average over the resulting decay rates, but this procedure is not amenable to the analytic analysis presented in this work. We have verified that both angular averaging methods yield similar results within 10\% accuracy.
\bibitem{coherent} In Fig.~\ref{Fig2} we removed coherent evolution due to the nonzero $J$ and $\delta h$. These coherent rotations are eliminated inherently with the $\pi_y$ control pulses considered in the rest of the paper.
\end{thebibliography}

\end{document}